\documentclass[%
 reprint,
 superscriptaddress,
 showpacs,
 amsmath,amssymb,
 pre
]{revtex4-1}
\usepackage{setspace}
\usepackage[export]{adjustbox}
\usepackage{lmodern}
\usepackage{siunitx}
\usepackage{etoolbox}
\usepackage{booktabs}
\usepackage{tabularx}
\usepackage[justification=raggedright]{caption}
\usepackage[justification=raggedright]{subcaption}
\usepackage{makecell}
\usepackage{graphicx}
\usepackage{dcolumn}
\usepackage{bm}
\usepackage{bbm}
\usepackage{bbold}
\usepackage{physics}
\usepackage{enumerate}
\usepackage[colorlinks=true,
		urlcolor=blue,
            linkcolor=blue,
            anchorcolor=blue,
            citecolor=blue
            ]{hyperref}

\begin{document}

\preprint{APS/123-QED}

\title{Instability in the quantum restart problem
}

\author{Ruoyu Yin}
\author{Qingyuan Wang}
\author{Eli Barkai}
\affiliation{Department of Physics, 
Institute of Nanotechnology and Advanced Materials, 
Bar Ilan University, Ramat-Gan 52900, Israel
}%

\date{\today}

\begin{abstract}
Repeatedly-monitored quantum walks with a rate $1/\tau$ 
yield discrete-time trajectories which are inherently random. 
With these paths the first-hitting time with sharp restart is studied. 
We find an instability in the optimal mean hitting time, 
which is not found in the corresponding classical random walk process.
%
This instability implies that a small change in parameters 
can lead to a rather large change of the optimal restart time.
We show that the optimal restart time versus $\tau$, 
as a control parameter, 
exhibits sets of staircases and plunges. 
The plunges, are due to the mentioned instability, 
which in turn is related to the quantum oscillations
of the first-hitting time probability, in the absence of restarts. 
Furthermore, 
we prove that there are only two patterns of staircase structures, 
dependent on the parity of the distance between the target and the source 
in units of lattice constant.
The global minimum of the hitting time,
is controlled not only by the restart time, as in classical problems, 
but also by the sampling time $\tau$. 
We provide numerical evidence that 
this global minimum occurs for the $\tau$ minimizing the mean hitting time, given restarts taking place after each measurement.
Last but not least, 
we numerically show that the instability found in this work is relatively robust against stochastic perturbations in the sampling time $\tau$.
\end{abstract}

\maketitle
\onecolumngrid

\section{Introduction}
Random search processes might be sent by chance to an undesired course away 
from a preset target position or area of space.
In such cases, the strategy of restart can be employed to tackle 
the decision-making conundrum of continuation or abortion of the process.
This idea is modeled as stochastic processes under restart
\cite{Gupta2022,Evans2020}.
Over the past decade,
the introduction of restart has opened a rapidly-expanding research field,
and related topics including the optimization of first-passage time,
non-equilibrium steady states, etc., 
have further propelled the expansion of this field 
\cite{Luby1993,Majumdar2011,Denis2014,Gupta2014,Pal2016,Eule2016,
Reuveni2016,restart,Majumdar2017,Belan2018,Igor2018,Evans2018,Boyer2019,
Igor2019b,Lukasz2019,Majumdar2020,Majumdar2020a,Evans2020,
Redner2020,Tal2020,Bressloff2020,Daniel2021,Majumdar2021,Ralf2021a,Ralf2021,Eliazar2021,Gupta2022,Blumer2022,Ahmad2022,Ray2022,Eliazar2023,Barkai2023}.

A general motivation for considering quantum dynamics with restart 
\cite{Majumdar2018,Rose2018,Belan2020,riera2020,Perfetto2021,Perfetto2021a,
Turkeshi2021,Haack2021,Magoni2022,Dattagupta2022,Sevilla2023,Anish2023}, 
are search processes using quantum computers (e.g. see \cite{Sabine2022,Wang2023}).
More specifically, the first-passage time in random walks,
defined as the time it takes a random walker to reach a target/threshold for the first time
\cite{Redner2001},
characterizes the efficiency of a classical search. 
Probably the most studied examples are the first-passage time 
of a Brownian motion or a random walker on the line.
For unbiased random walk in unbounded space,
with restarts, namely resetting a random walker to its initial site at some time $t_r$, 
the expected first-passage time exhibits one unique minimum \cite{Majumdar2011,Gupta2022}.
As conveyed by the authors' previous publication (see Ref. \cite{Ruoyu2023}),
completely different behaviors are found in the quantum world within the context of {\em deterministic} restarts (see definitions below).
Employing repeated measurements/monitoring with a period $\tau$ at the preset target state, one can define the {\em quantum hitting time}, as the quantum counterpart of first-passage time, via the record of measurement outcomes (see details below).
%
Using deterministic restarts,
there appear multiple minima, 
instead of a sole minimum, of the mean hitting time under restart
\cite{Ruoyu2023}.
%
%
More remarkable is the periodical staircase structure of the optimal restart time,
along with plunges/rises, which manifest a kind of instability, 
as a quantum signature in the restart framework.
This is attributed to the existence of several minima in the expected hitting time.
In Ref. \cite{Ruoyu2023} we treated this staircase structure, 
but only when the target is set at the initial state, i.e. the return case.
Here in this paper, we will delve into the universality of staircase structures 
and accompanied instability for general choices of the target location.
%
%
More profoundly, the staircase structure converges to a unique structure 
in the limit of rare measurements, namely large $\tau$ limit, depending only on the parity of the distance 
between the initial state and the target.

The process of restarts for monitored quantum walks can be implemented on quantum computers \cite{Yin2023c}.  
One may wonder, 
given the instabilities we find, 
whether small stochastic perturbations 
in the sampling times, 
lead to the wipe-out of novel quantum features. 
When does the noise destroy the quantum effects discussed in this paper? 
And does a small amount of noise severely affect our results?
We will answer these queries towards the end of this paper.

Another natural question is how to find the global optimum 
with respect both to the restart time and the measurement period. 
Both these parameters are, at least in principle, control parameters, 
in the mentioned quantum computer experiments. 
In the classical world, 
the sampling time is of little concern, 
and one may try to detect the walker continuously. 
In the quantum case, 
due to the Zeno effect 
which typically inhibits detection, 
sampling cannot be performed continuously, 
and hence the sampling time becomes an important parameter.
Therefore, in the quantum case, 
we have two control parameters, 
while typically in the classical restart problem, we minimize hitting times by changing the restart time only.
With restarts the fast sampling leads to Zeno physics and very long mean hitting times, 
when the initial and target states are non-identical.
So clearly too small $\tau$ or too large $\tau$ are not optimal for the sake of fast search, 
and similarly with respect to the restart time.
The analysis for this global optimization problem will be treated in the final stage of this paper.
This paper is structured as follows: 
we present in the first place the concept of quantum hitting time,
or the first-detected-passage time (in the absence of restart),
for a periodically-monitored quantum walk (Sec. \ref{FHT}) 
\cite{Krovi2006,Gruenbaum2013,Dhar2015,Dhar2015a,Friedman2017a,Thiel2018a,Thiel2018,yin2019,Ruoyu2020}.
Then the mean hitting time with restart is provided in Sec. \ref{FHTr},
for the model of a tight-binding quantum walk on an infinite line.
Further analysis for the optimization problem of the optimal restart time, exposing instabilities in the quantum restart problem,
expounds the main results of the paper in Sec. \ref{optMHT}.
We check the robustness of our results to noise in Sec. \ref{robust}. 
The minimization of the mean hitting time, 
with respect to both the measurement period and the restart time, 
is studied in Sec. \ref{DoubleOpt},
namely we search for the global minimum of the mean hitting time.
%
We close the paper with a summary and discussions.
A brief summary of part of our results was recently presented in a Letter 
\cite{Ruoyu2023}.

\section{The quantum first-hitting time in absence of restart}
\label{FHT}
We first introduce the quantum first-hitting time problem,
which is based on the continuous-time quantum walk \cite{Farhi1998},
but with repeated monitoring (measurements).
The Hamiltonian we will employ in this paper is a tight-binding model of a single particle, sometimes called the walker, on an infinite line
\begin{equation}\label{ham}
    H = -\gamma \sum_{x=-\infty}^\infty
        \left[
              \ket{x}\bra{x+1}+\ket{x+1}\bra{x}
        \right]
        .
\end{equation}
Here $\gamma$ is the hopping rate and set as $1$ in what follows.
This is a lattice walk \cite{Blumen2011}, 
as the particle can occupy the integers denoted with the ket $\ket{x}$,
and the hopping is to nearest neighbors only. 
The energy spectrum is $E(k)=-2\cos(k)$ in units of $\gamma$,
and the eigenfunctions of $H$ are 
$\psi_k(x)=e^{ikx}/\sqrt{2\pi}$ with $k\in [-\pi,\pi]$.
Such tight-binding Hamiltonians are used extensively in condensed matter. 
Then the propagation of the quantum wave packet (in the absence of measurement)
is described by the probability of finding the particle 
at state $\ket{x_\text{d}}$ starting from $\ket{x_0}$ (both are spatial states of the lattice),
i.e.
\begin{equation}\label{prop}
    P(x_\text{d},x_0,t) =\left| \bra{x_\text{d}}\ket{\psi(t)} \right|^2
                = \left| \bra{x_\text{d}}\hat{U}(t)\ket{x_0} \right|^2
                = \left| i^{|x_\text{d}-x_0|} J_{|x_\text{d}-x_0|}(2t) \right|^2
                = J^2_{|x_\text{d}-x_0|} (2t),
\end{equation}
where $\ket{\psi(t)}$ is the solution to the Schrödinger equation for the Hamiltonian Eq. (\ref{ham}),
$\hat{U}(t)=e^{-iHt}$ is the unitary operator with $\hbar$ set as $1$ in what follows,
and $J_n(x)$ is the Bessel function of the first kind.
This is led by the cosine law of the dispersion relation $E(k)$
\cite{Friedman2017a,Thiel2018a}.
Hence the quantum walker's travel is ballistic \cite{Grover1971,Norio2005,Yaron2008,Hoyer2010},
and vastly distinct from the Gaussian spreading of a classical walker on a similar lattice 
\cite{Redner2001}.

\begin{figure}[ht]{}
\centering
\includegraphics[width=0.5\linewidth]{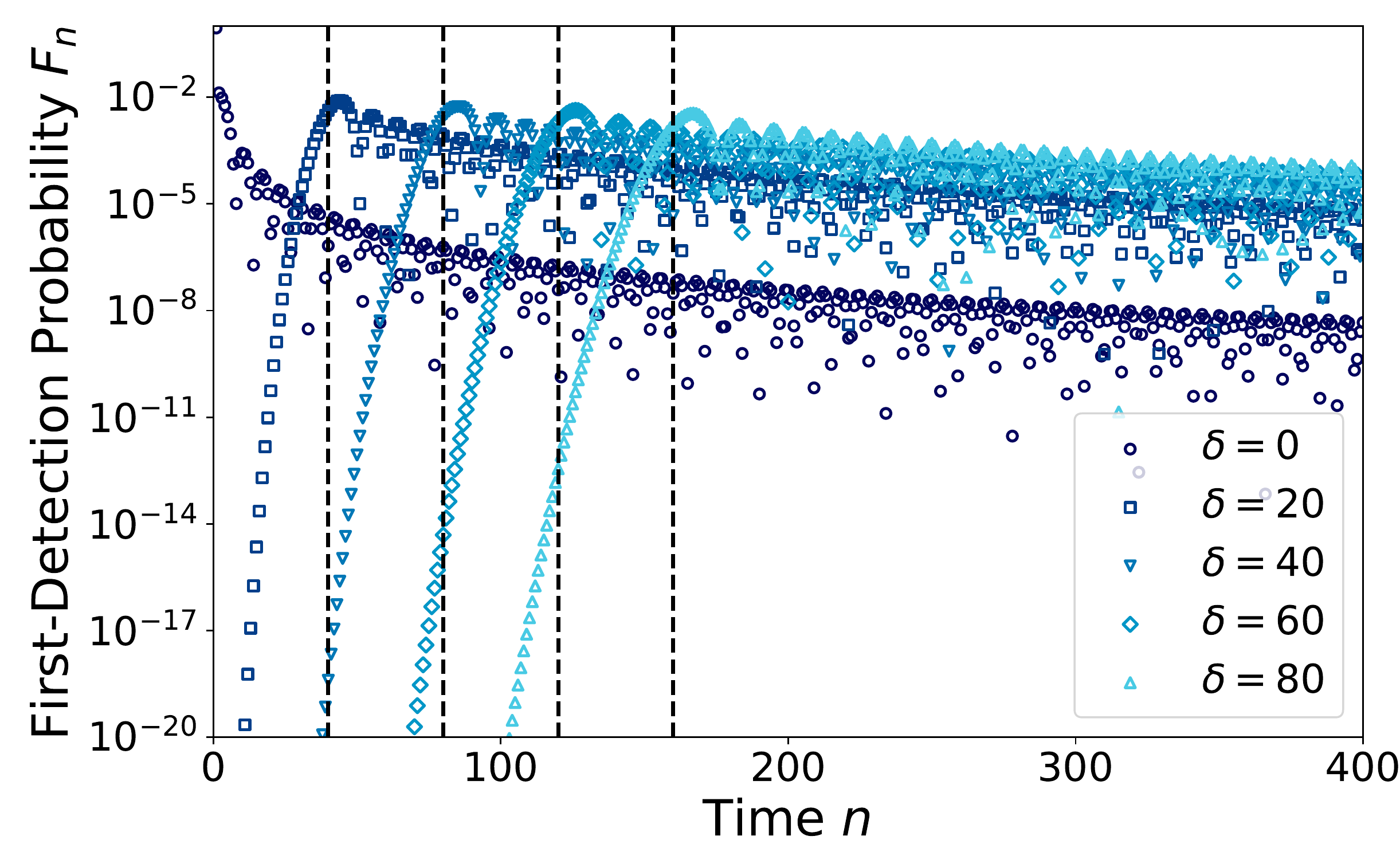}
\caption{$F_n$ vs. $n$ for different $\delta$, with $\tau=0.25$ ($\gamma=1$ as mentioned).
The dashed lines represent the $n_\text{inc}=\delta/2\tau$, 
close to the $\max(F_n)$.
For a thorough discussion on the quantum first-hitting probability,  
see Ref. \cite{Thiel2018a}.
}
\label{fnn}
\end{figure}
To determine the time when the particle arrives at some target for the first time,
one cannot simply observe the “walking” process of a quantum particle,
since according to Born's rule 
it ``freezes'' the particle at some eigenstate of the observable.
This gives rise to a fundamental process and debatable problem 
known as time-of-arrival (TOA) in quantum physics 
\cite{Muga2000}.
A way to solve this issue renders 
the framework of “monitoring” the (unitary) walking process built 
to define the quantum {\em first-detected-passage} time or first-hitting time,
via some measurement protocol,
within which the stroboscopic measurement protocol has been investigated in full detail
\cite{Krovi2006,Gruenbaum2013,Dhar2015,Dhar2015a,Friedman2017a,
Thiel2018a,Thiel2018,yin2019,Ruoyu2020,Sabine2022}.
The stroboscopic monitoring protocol states the following:
a quantum walker is initially dispatched 
at a localized state $\ket{x_\text{0}}$ (source), 
and one attempts to measure the walker on the detected state $\ket{x_\text{d}}$ (target) 
at fixed times, $(\tau, 2\tau, 3\tau,\cdots)$.
In between the measurement attempts,
the system undergoes free evolution dictated by the Schrödinger equation.
We employ von Neumann (strong) measurement described by the projection 
$\hat{D}=\ket{x_\text{d}}\bra{x_\text{d}}$,
so the outcome of each measurement is yes (detection) or no (null detection)
with probabilities determined by the Born rule.
For the {\em first} hitting of the walker at the $n$th measurement attempt,
the output of the experiment must be “no, no, no, no, ..., yes”,
namely a final success at the $n$th attempt following previous $n-1$ failure,
since the experiment is done once the walker hits the detector,
i.e. a yes event is recorded.
Each null detection acts as a wipe-out of the component of the wave function at the target
\cite{Gruenbaum2013,Dhar2015a,Friedman2017a}.
We will explain the effects from the null detection below,
which manifest themselves in the quantum renewal equation.
Inherently, $n$ is a random variable, 
and is defined as the first-detected-passage time or hitting time (in units of $\tau$). 
We denote the amplitude of finding the walker for the {\em first} time
at the $n$th attempt by $\phi_n$ \cite{Dhar2015a}.
Using the quantum renewal equation,
one can in principle solve for the quantum first-hitting amplitude
\cite{Friedman2017a}:
%
\begin{equation}\label{qc}
  \begin{aligned}
    \phi_n (x_\text{d},x_0) =   \bra{x_\text{d}}\hat{U}(n\tau)\ket{x_0} 
                                - \sum_{m=1}^{n-1} 
                                \bra{x_\text{d}} 
                                \hat{U}[(n-m)\tau] 
                                \ket{x_\text{d}} 
                                \phi_m .
  \end{aligned}
\end{equation}
%
Technically one uses Eq. (\ref{prop}) and iterations to solve this equation 
or one may employ generating function techniques. 
For example, $\phi_1 = i^{|x_\text{d}-x_0|} J_{|x_\text{d}-x_0|}(2\tau)$, 
which is expected from basic quantum mechanics Eq. (\ref{prop}), 
while $\phi_2 = i^{|x_\text{d}-x_0|}
\left[J_{|x_\text{d}-x_0|}(4\tau) - J_{0}(2\tau)J_{|x_\text{d}-x_0|}(2\tau) \right]$.
Eq. (\ref{qc}) is a quantum counterpart of the well-known classical renewal equation, 
which is discussed in Ref. \cite{Redner2001}. 
From Eq. (\ref{qc}), one can readily find the essential effects 
from the repeated local measurement conditioned with null outcome:
the first-hitting amplitude is indeed related to 
the measurement-free transition amplitude 
$\bra{x_\text{d}}\hat{U}(n\tau)\ket{x_0}$,
but the measurement-free return amplitude 
propagated from the {\em prior} first-hitting amplitude,
$\bra{x_\text{d}}\hat{U}[(n-m)\tau]\ket{x_\text{d}}\phi_m$ ($m<n$), 
should be subtracted, 
since they represent the events that have been aborted by the null detection 
from the statistical ensemble.

Using Eq. (\ref{qc}), one then finds $F_n$, the first-hitting probability at time $n$,
\begin{equation}
    F_n = |\phi_n|^2.
\end{equation}
%
With Eq. (\ref{prop}) and $\delta := |x_\text{d}-x_\text{0}|$, 
that denotes the distance between the source and target 
in units of the lattice constant, 
one obtains, via iterations \cite{Friedman2017a}:
%
%
\begin{equation}\label{eq1a}
\begin{aligned}
  F_1 &= |\phi_1|^2 = J^2_\delta(2\tau), \\
  F_2 &= 
        [J_\delta(4\tau) - J_0(2\tau)J_\delta(2\tau)]^2, \\
  F_3 &= 
        [J_\delta(6\tau) - J_0(4\tau)J_\delta(2\tau) 
         -J_0(2\tau)J_\delta(4\tau) + J^2_0(2\tau)J_\delta(2\tau)]^2, \\
      &\;\;\vdots
\end{aligned}
\end{equation}
A numerical demonstration is presented in Fig. \ref{fnn} for different values of $\delta$. 
As one may witness in the figure, 
besides the oscillatory decays,
the maximum of $F_n$ plays the role of a transition point 
distinguishing a rapid growth and a slow decay of $F_{n}$.
The dashed lines at $n=40, 80, 120, 160$ 
(correspond to $\delta=20, 40, 60, 80$)
approximately point to the maxima,
and those special values of $n$ are given by the maximal group velocity 
$\max(v_g)=\max\left[|\partial_k E(k)|\right]=2$, the distance $\delta$,
and the measurement period $\tau$ in a kinematic fashion:
$n_\text{inc}:=\delta/2\tau$, 
hence we regard it as the ``incidence'' time of the ballistic-propagating wave-front
\cite{Thiel2018a}.
Around the $n_\text{inc}$,
the chance for the detector to be hit reaches the maximum.
Detailed analysis of $F_n$ is provided in \cite{Friedman2017a,Thiel2018a}.
For a classical random walk, roughly speaking one has a peak 
in $F_n$ that is determined by diffusive motion 
and the initial distance of target and source, 
followed by a power-law decay. 
In the quantum world non-trivial oscillations determined by a phase, superimpose on power-law decay are found \cite{Thiel2018a}. 
Yet another difference between the classical and quantum hitting time problem 
is that the former is recurrent, while the latter is not, 
namely in general $\sum_{n=1}^\infty F_n<1$ in the quantum case
\cite{Krovi2006,Thiel2020,Thiel2020D}.

Now we will incorporate the restart framework with the quantum hitting time.
As shown before, the ballistic propagation is a quantum advantage in faster search
(over classical diffusive motion),
however, as mentioned, $F_n$ is unfortunately non-normalized in this $1$D model 
(also in many finite systems),
leading to infinite mean fitting times \cite{Krovi2006}.
This indicates that the probabilistic nature of quantum dynamics 
sends the walker to undesired ``trajectories'' far away from the target,
or to the states orthogonal to the detected state in the Hilbert space {\em forever}
\cite{Thiel2020D}.
A systematic strategy to solve this problem, 
inspired from processes happening in nature
\cite{Bel2009a,Bel2009,MMOptRestart}
or algorithmic methods used in classical computers \cite{Gomes1998,Gomes1998b},
is to perform restarts to rescue a process that probably enters a wrong track.
We expect the approach of restart to take advantage of the ballistic spreading
of the wave-front in each single {\em run}
(one {\em run} is a monitored process between restarts),
and in the mean time, to guarantee the detection of the quantum walker 
(see Appendix \ref{appdix1}),
so that the quantum advantage of faster search 
is reinforced to pronounce the supremacy of the quantum search.

\section{The first-hitting statistics under restart}
\label{FHTr}
%
We will consider the deterministic restart (or sharp restart) strategy \cite{Pal2016},
namely, after every $r$ failed attempts in detecting the particle, 
the system is reset to the initial state to restart the monitored process. 
This strategy has been proven as the outperforming one,
in the sense of unfailingly achieving the lowest minimum of the mean hitting time
among all random restart strategies 
\cite{Luby1993,restart}.
Let $n_R$ be the first-hitting time under restart in units of $\tau$. 
The general formula for the mean hitting time under sharp restart 
$\expval{n_R(r)}$ (the variable $r$ means $r$ steps between restarts) is
\begin{equation}\label{eq9}
\expval{n_R(r)} = r {1-P_{\rm det}^r \over P_{\rm det}^r}  + \expval{n}^r_\text{cond},
\end{equation}
%
where $P_{\rm det}^r := \sum_{n=1}^r F_n$ 
is the detection probability within $r$ attempts, and 
\begin{equation}
    \expval{n}^r_\text{cond}:= 
            {
            \sum_{n=1}^r nF_n 
            \over 
            P_{\rm det}^r
            },
\end{equation}
which computes the conditional mean 
of the first-hitting provided the particle is detected within $r$ attempts.
The $F_n$ are the restart-free probabilities given with Eq. (\ref{eq1a}).
This result has been presented in Refs. \cite{Luby1993,Pal2021},
and we also provide an alternative derivation in Appendix \ref{appdix2}.
We note that in the large $r$ limit, 
for classical random walks in dimension one,
$P_{\rm det}^r\to 1$, 
then $\expval{n_R(r)} \to \expval{n}^r_\text{cond}$,
while for the quantum walk Eq. (\ref{ham}),
as mentioned above,
$P_{\rm det}^r\to \sum_{n=1}^\infty F_n<1$. 
This implies that the first term in the right-hand side of Eq. (\ref{eq9}),
$r (1-P_{\rm det}^r)/P_{\rm det}^r$, cannot be neglected in the large $r$ limit.
We will show later that for the quantum walk on an infinite line, 
$\expval{n_R(r)} \sim a r + b$ when $r\to\infty$, 
with $a=(1-P_{\rm det}^{r=\infty})/P_{\rm det}^{r=\infty}$, 
$b=\expval{n}^{r=\infty}_\text{cond}$,
namely, $\expval{n}^r_\text{cond}$ in the quantum case converges to a finite number.
%
\begin{figure*}[ht]{}
\centering
\includegraphics[width=0.55\linewidth]{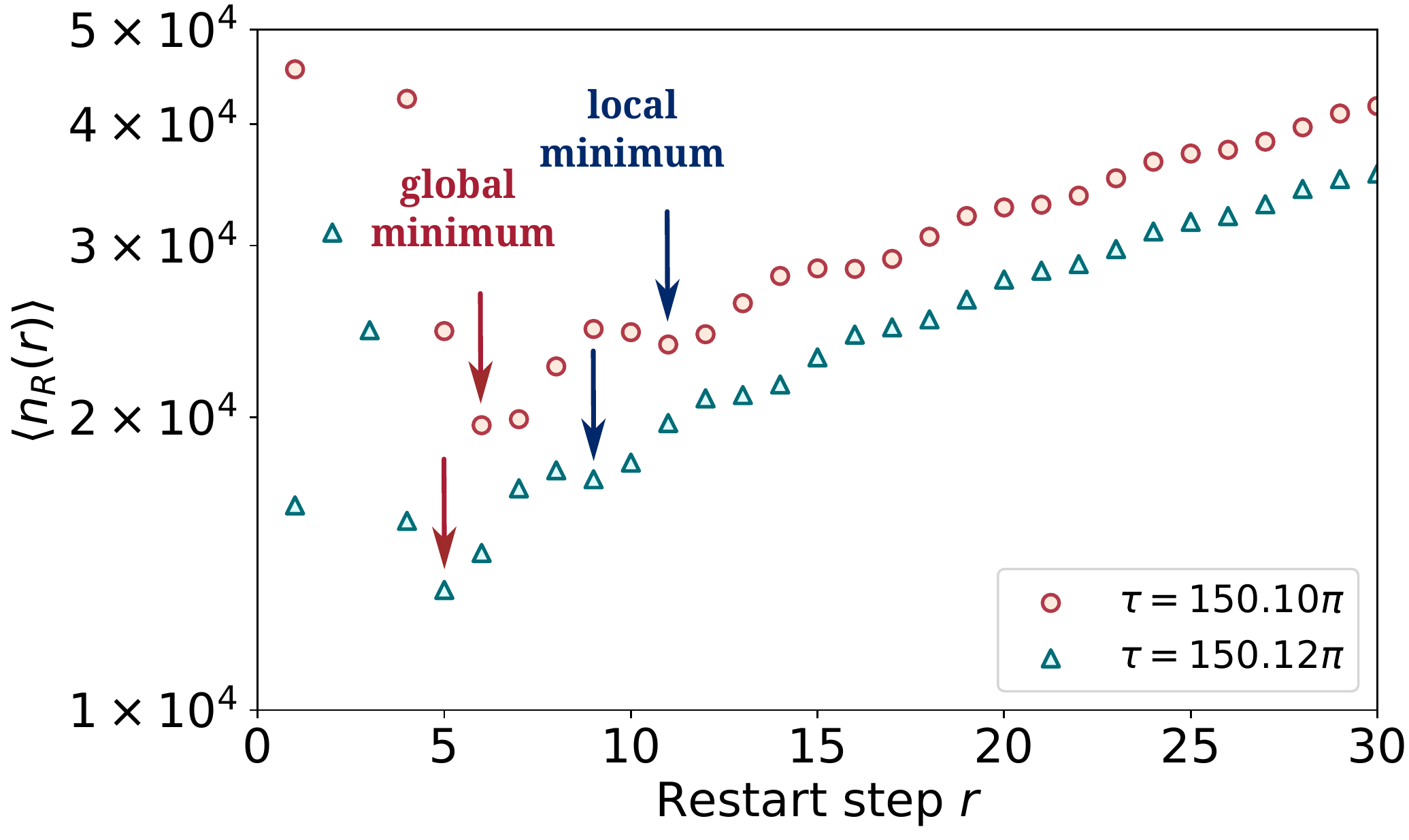}
\caption{
$\expval{n_R(r)}$ vs. $r$ for $\delta=60$, with different $\tau$.
We see the oscillations of $\expval{n_R(r)}$ render the presence of several extrema (see the arrows),
and the global minimum is located in small $r$,
which will be analyzed below.
When $\tau$ changes from $150.10\pi$ to $150.12\pi$,
we see a big change in the minimum of $\expval{n_R(r)}$, 
namely $\min{\expval{n_R(r)}}=19615\to13287$,
with the optimal restart step changing from $6$ to $5$.
This shows that a slight change of $\tau$ causes a very large change of the optimum,
indicating a type of instability.
}
\label{rmht}
\end{figure*}
%
%
%

In principle, substituting the first-hitting probability $F_n$ Eq. (\ref{eq1a}) 
into the formula for the mean detection time under restart Eq. (\ref{eq9}), 
we obtain $\expval{n_R(r)}$ for this specific model,
an approach that is used to test the approximations studied below, 
the latter providing insights into the behaviors of the quantum restart.
We provide a numerical demonstration of the general landscape of $\expval{n_R(r)}$ in Fig. \ref{rmht}.
The first remarkable feature is the oscillations of $\expval{n_R(r)}$, 
leading to multiple extrema, 
which is vastly different from the classical restart with one distinct minimum
\cite{Gupta2022}.
Second, we notice that a small variation of the sampling time $\tau$
leads to a large change of the optimum,
which switches in this example from 
$\expval{n_R(6)}=19615$ to $\expval{n_R(5)}=13287$.
This suggests a type of instability as mentioned in the introduction.
Note the measurement periods $\tau$'s are chosen large here 
($\tau=150.10\pi, \,150.12\pi $), 
which allows the application of asymptotic methods on our problem
soon to be discussed.

More precisely, we list a few analytical expressions for $\expval{n_R(r)}$
using Eqs. (\ref{qc},\ref{eq1a},\ref{eq9}),
%
\begin{equation}\label{meaneg}
\begin{aligned}
    \expval{n_R(1)} &= {1 \over F_1} = {1 \over J^2_\delta(2\tau)}, 
    \quad\quad\quad\,\,\, \\
    \expval{n_R(2)} &= {2 - F_1 \over F_1 + F_2} 
    = {2- J^2_\delta(2\tau) \over J^2_\delta(2\tau) 
    + [J_\delta(4\tau) - J_0(2\tau)J_\delta(2\tau)]^2}, \\
    \expval{n_R(3)} &= {3-2F_1-F_2 \over F_1+F_2+F_3} 
    = {3-2J^2_\delta(2\tau)-[J_\delta(4\tau) - J_0(2\tau)J_\delta(2\tau)]^2 
    \over F_1+F_2+F_3}, \\
    \expval{n_R(4)} &= {4-3F_1-2F_2-F_3 \over F_1+F_2+F_3+F_4}, \;\;
    \expval{n_R(5)} = {5-4F_1-3F_2-2F_3-F_4 \over F_1+F_2+F_3+F_4+F_5}, 
    \;\;\cdots
\end{aligned}
\end{equation}
%
Those expressions are cumbersome, 
however we will analyze particular limits where we may provide insights, 
e.g. large measurement period $\tau$. 
In what follows, 
our discussions will be based on choosing certain values for $\delta$ 
to gain some insights (e.g. $\delta=0,1,2,\cdots$),
and investigating how to optimize the mean hitting time under restart.
One of our goals is to find the optimal restart time $r$, denoted by $r^*$, 
which will be in general a function of $\tau$ and $\delta$. 
The corresponding $\expval{n_R(r^*)}$ is then the optimal in the mean sense.

\section{Optimization of the mean hitting time}\label{optMHT}
\subsection{Return case: $\delta=0$}
We start with the ``return'' case where the detector is put at the origin 
to monitor the walker's first return, 
namely, $\delta=0$ \cite{Friedman2017a}.
We first consider the Zeno limit $\tau\to 0$.
Specially, in this limit,
we have the following asymptotics for $\expval{n_R(r)}$ using Eq. (\ref{meaneg}):
\begin{equation}
    \expval{n_R(1)} \sim 1+2\tau^2,\;\;\; \expval{n_R(2)} \sim 1+4\tau^2,\;\;\;
    \expval{n_R(3)} \sim 1+6\tau^2, \cdots 
\end{equation}
Clearly, the minimum is $\expval{n_R(1)}$.
Hence the optimal restart $r^* =1$ in the Zeno regime,
and this is also intuitive, 
since the wave function is nearly frozen at the origin in the Zeno limit 
\cite{Misra1977},
and it is always the best choice to restart after each failed measurement. 
%
\begin{figure}[htbp]{}
\centering
\includegraphics[width=0.45\linewidth]{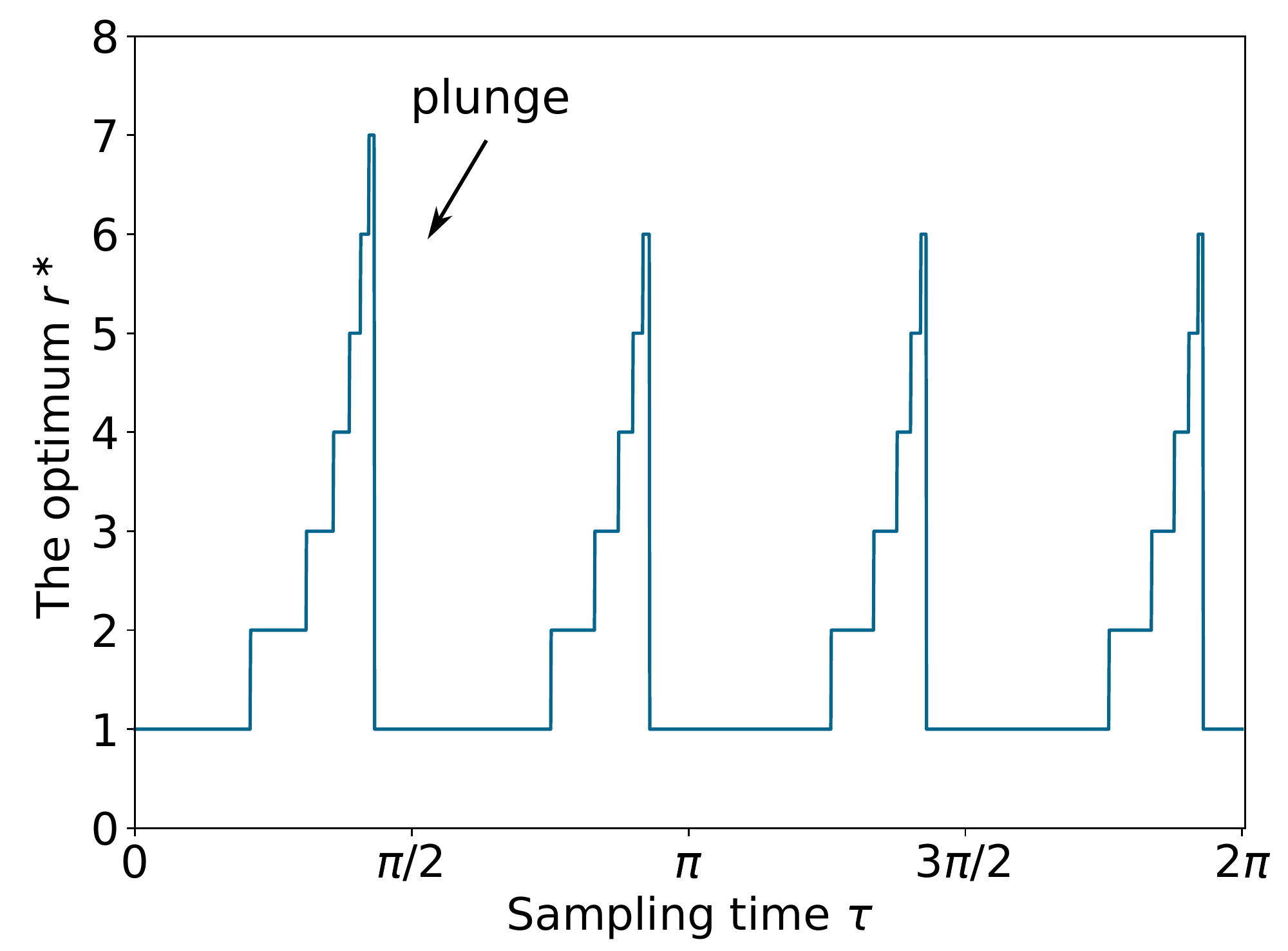}
\caption{The optimal restart time $r^*$ as a function of $\tau$ for $\delta=0$.
We see the staircase structure accompanied by periodical plunges (see the arrow).
As we increase $\tau$ we witness a convergence of the staircase structure.
}
\label{Bess0}
\end{figure}
\begin{figure}[ht]{}
\centering
\includegraphics[width=0.45\linewidth]{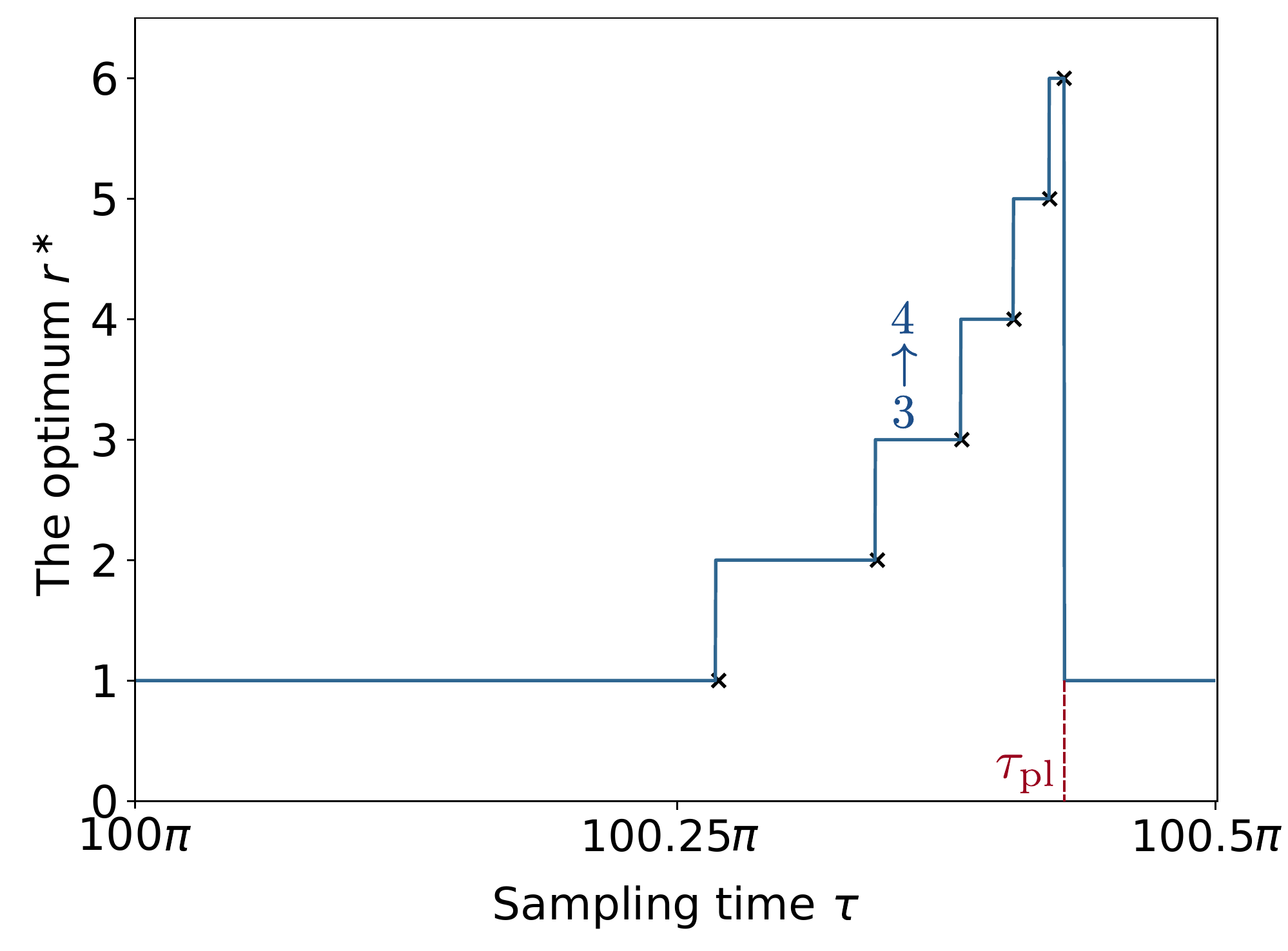}
\caption{$r^*$ versus $\tau$ in the large $\tau$ limit 
exhibits a limiting staircase and plunge. 
Here we use $\tau\in[100\pi, 100.5\pi]$ and $\delta=0$.
The cyan lines are the exact results, 
and the black crosses represent the approximations using Table \ref{trantau}.
$\tau_\text{pl}$ here is approximated as $k\pi/2+1.353$ with $k=200$.
}
\label{ore0}
\end{figure}
\begin{figure}[htbp]{}
\centering
\includegraphics[width=0.45\linewidth]{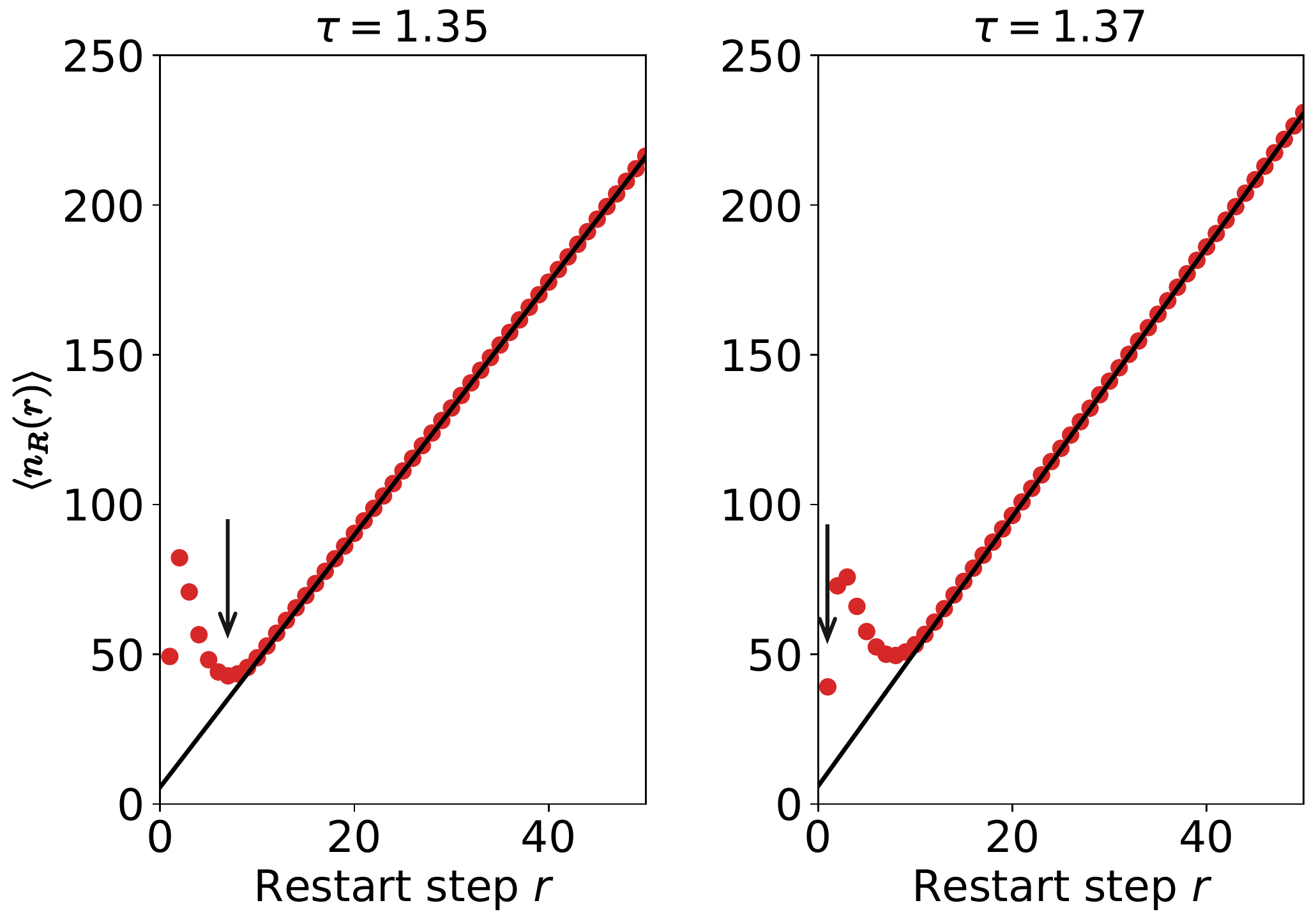}
\caption{$\expval{n_R(r)}$ vs. $r$ 
in the vicinity of the plunge $\tau$ ($\approx1.36$), 
see the drop of $r^*$ ($7\to1$) 
at the leftmost staircase in Fig. \ref{Bess0}.
There are two minima competing with each other, 
and a small change of $\tau$ ($1.35 \to 1.37$) results in different optima.
On the left panel $r^*=7$ and on the right $r^*= 1$ 
(see the arrows).
Thus a small change of $\tau$ creates a plunge of $r^*$. 
The solid lines represent the approximations to $\expval{n_R(r)}$ 
in large $r$ limit,
i.e. $\expval{n_R(r)} \sim r (1-P_{\rm det}^{r=\infty})/P_{\rm det}^{r=\infty}+\expval{n_R(\infty)}_\text{cond}$,
where $\expval{n_R(\infty)}_\text{cond}= \sum^\infty_{n=1} n F_{n}/\sum^\infty_{n=1} F_{n}$.
}
\label{tran}
\end{figure}
%

For large $\tau$,
with the large $x$ asymptotics for $J_n(x)$,
namely, $J_n(x) \sim \sqrt{2/\pi x}\cos(x-\pi n/2-\pi/4)$ \cite{Math2007},
we can re-express $F_n$ for finite $n$ as 
\cite{Friedman2017a}
\begin{equation}\label{fna}
    F_n(\tau) \sim {1 \over n\pi \tau} 
                    \cos^2 
                    \left( 2n\tau-{\pi \over 4} \right).
\end{equation}
%
In this limit there is a simple relation between the probability of first-hitting time 
and the measurement-free wave function. 
Namely, $F_n \sim P(0,0,t)=|\bra{0}\ket{\psi(t)}|^2=J_0^2(2t)$ with $t=n \tau$ 
and $\ket{\psi(t)}$ is the solution of the Schrödinger equation,
in the absence of measurement, see Eq. (\ref{prop}). 
%
%
If $2\tau$ is a multiple of $\pi$, we find using Eq. (\ref{fna})
\begin{equation}\label{fn2p}
    F_n \sim {1 \over 2 n \pi \tau}.
\end{equation}
Thus, the restart-free $F_n$ decays monotonically with $n$.
For such a case, the best strategy of restart 
is to use $r=1$, namely, to restart after each measurement. 
We then have 
\begin{equation}\label{eqadded}
    r^* = 1, \quad \text{when} \,\,\, 2\tau = k\pi,
\end{equation}
where $k$ is an integer.
Using Eqs. (\ref{meaneg},\ref{fn2p}), 
$\expval{n_R(r^*)} \sim 2 \pi \tau$. 

We now study how this optimal restart changes 
when we vary the sampling time $\tau$, 
in particular, what is $r^*$ as a function of $\tau$?
We have found that $r^*$ exhibits a set of staircases, 
see Fig. \ref{Bess0}.
Clearly this behavior is far from the classical limit, 
and it is due to the oscillations of the first-hitting probability $F_n$. 
The optimal choice of $r$, 
presented in Fig. \ref{Bess0},
is a periodical function of $\tau$, 
with a period of $\pi/2$. 
Starting with $\tau=\pi k/2 \gg 1$, where $k$ is an integer, 
when we increase $\tau$ slightly beyond this critical point,
we find that $\expval{n_R(1)}$ is the optimal, 
i.e. the fastest approach to restart is still $r^*=1$. 
We see transitions as $\tau$ is varied, from $r^* =1$ to $r^*=2$, and then to $r^*=3$, etc.,
and finally a plunge to $r^*=1$, and this is repeated.
The staircase for small $\tau$ is not identical to that for large $\tau$, 
however, as to be shown below in the latter limit, we will reach a particular pattern of staircases,
presented in Fig. \ref{ore0}.

Now we also use the fact that $F_n \ll 1$ in the limit under study, 
namely the large $\tau$ limit. 
Using Eq. (\ref{meaneg}), we find that the condition for $\tau$, 
for the $r^*=1$ to $r^*=2$ transition, i.e. $\expval{n_R(1)}=\expval{n_R(2)}$,
reads
\begin{equation}\label{f2}
    F_1(\tau) = F_2(\tau),
\end{equation}
and $F_n$ are given in Eq. (\ref{fna}).
We then find a sequence of transitions,
and due to the fact that $F_n$ is small, 
we find using Eq. (\ref{meaneg}), 
the transition $r^*=2 \to  r^*=3$ at $\tau$ which solves
\begin{equation}\label{f3}
    F_3(\tau) = [F_1(\tau) + F_2(\tau)]/2,
\end{equation}
and similarly for the $r^*=3 \to r^*=4$ transition 
shown in Fig. \ref{ore0},
\begin{equation}\label{f4}
    F_4(\tau) = [F_1(\tau) + F_2(\tau) + F_3(\tau)]/3.
\end{equation}
We see that the transitions are taking place 
when $F_n$ is the mean of all the proceeding $F_i$'s. 
At some stage these equations cannot be solved,
in the sense that there is no $\tau$ giving a valid solution.  
We encountered already such a situation, 
and that is the Zeno limit. 
%
We can use Eq. (\ref{fna}) for $F_n(\tau)$ and find, 
with a simple computer program of calculation, 
the transitions in $r^*$ at special sampling times. 

Note that we have plunges where $r^*$ falls to the value $1$ 
(see the arrow in Fig. \ref{Bess0}).
In Fig. \ref{tran}, we choose as an example two values of $\tau$ in the vicinity of $\tau\approx 1.36$.
At this value we have a plunge (see Fig. \ref{Bess0}).
As shown in Fig. \ref{tran}, 
there are two minima competing with each other, 
and the global minimum switches between them when $\tau$ is slightly varied.
At the exact transition time $\tau$ for the plunge,
the two minima are identical. 
Thus the system exhibits an instability in the sense that 
small changes of $\tau$ create large difference 
in the optimal restart time $r^*$.

We now calculate the sampling time $\tau_\text{pl}$, where plunges are found. 
Let $\tau = k \pi/2 + \epsilon$ and $0<\epsilon<\pi/2$. 
As mentioned in Eq. (\ref{eqadded}), if $\epsilon=0$, $r^*=1$. 
We then denote $\epsilon_{1\to2}$ as the value of $\epsilon$ 
where we have a transition from $r^*=1$ to $r^*=2$, 
similarly for other transitions.
In between the transition $\epsilon$,
namely for each interval $[\epsilon_{k\to k+1}, \epsilon_{k+1\to k+2}]$,
we will check whether $\expval{n_R(k+1)}$ remains the minimum, 
and especially compare it with $\expval{n_R(1)}$ 
in case we miss the plunge to $r^*=1$.
With Eq. (\ref{fna}) and Eqs. (\ref{f2}-\ref{f4}) 
we get Table \ref{trantau} which gives the values or $\epsilon$ for the various transitions,
and check the minimum in Table \ref{trantau0}. 
\newcommand\tstrut{\rule{0pt}{2.4ex}}
\newcommand\bstrut{\rule[-1.0ex]{0pt}{0pt}}
\begin{table*}[ht]
\centering
\caption{The $\epsilon$ at the transitions of $r^*$ for $\delta=0$.}
\label{ttau}
\subcaption{The values of $\epsilon_{k\to k+1}$.}
\begin{tabular}{ cccccccccccccc } 
 \Xhline{2\arrayrulewidth}
 \multicolumn{2}{c}{$\epsilon_{k\to k+1}$} & 
 \multicolumn{2}{c}{$\epsilon_{1\to2}$}  & 
 \multicolumn{2}{c}{$\epsilon_{2\to3}$}  & 
 \multicolumn{2}{c}{$\epsilon_{3\to4}$}  &
 \multicolumn{2}{c}{$\epsilon_{4\to5}$}  &
 \multicolumn{2}{c}{$\epsilon_{5\to6}$}  &
 \multicolumn{2}{c}{$\epsilon_\text{pl}$}  \\
 \multicolumn{2}{c}{Value} & \multicolumn{2}{c}{0.850} & 
 \multicolumn{2}{c}{1.081} & \multicolumn{2}{c}{1.204} & 
 \multicolumn{2}{c}{1.280} & \multicolumn{2}{c}{1.332} & 
 \multicolumn{2}{c}{1.353} \\
 %
 \Xhline{2\arrayrulewidth}
\end{tabular}
\label{trantau}
\bigskip
\subcaption{The relation between $\expval{n_R(k)}$ and $\expval{n_R(1)}$ in $[\epsilon_{k-1\to k}, \epsilon_{k\to k+1}]$.}
\begin{tabular}{ c|c|c|c|c|c } 
\Xhline{2\arrayrulewidth}
 {$[\epsilon_{1\to2}, \epsilon_{2\to3}]$} & 
 {$[\epsilon_{2\to3}, \epsilon_{3\to4}]$} & 
 {$[\epsilon_{3\to4}, \epsilon_{4\to5}]$} & 
 {$[\epsilon_{4\to5}, \epsilon_{5\to6}]$} & 
 {$[\epsilon_{5\to6}, \epsilon_\text{pl}]$} & 
 {$[\epsilon_\text{pl}, \pi/2]$} \tstrut \bstrut \\
%
%
 {$\expval{n_R(2)} < \expval{n_R(1)}$} & 
 {$\expval{n_R(3)} < \expval{n_R(1)}$} & 
 {$\expval{n_R(4)} < \expval{n_R(1)}$} & 
 {$\expval{n_R(5)} < \expval{n_R(1)}$} & 
 {$\expval{n_R(6)} < \expval{n_R(1)}$} & 
 {$\expval{n_R(1)} < \expval{n_R(6)}$} \tstrut \bstrut\\ 
\Xhline{2\arrayrulewidth}
\end{tabular}
\label{trantau0}
\end{table*}
%
%
Note that for the transitions with large $r^*$,
$\epsilon$ is accumulating close to $\pi/2$,
and hence the plateaus in optimal $r$ are very small.
Furthermore, when $1.353<\epsilon<\pi/2$, 
$r^*$ drops to $1$,
i.e. a sudden plunge as mentioned.
So we have
\begin{equation}
\tau_\text{pl}=1.353+\pi k/2
\end{equation} 
for large $k$,
and in other words, let $\tau_\text{pl}=\pi k/2+\epsilon_\text{pl}$,
then $\epsilon_\text{pl}=1.353$.
As shown in Fig. \ref{Bess0},
$r^*$ is plotted versus $\tau$, and we see staircases,
where $r^*$ is increasing by unit steps, as our theory predicts,
and then a sudden plunge.
The staircase structure and plunges of $r^*$ versus $\tau$ 
are found in the whole range of $\tau$, 
and are not limited to large $\tau$ limit studied
presented analytically in Table \ref{ttau}. 


A general formalism to obtain the transition $\epsilon$
is stated as follows.
We find using $F_n \ll 1$ valid from the large $\tau$ limit and Eq. (\ref{eq9}),
%
\begin{equation}\label{eq13}
    \expval{n_R(r)} \sim r \big/\sum_{n=1}^r F_n.
\end{equation}
Then the condition for the transition $\tau$ 
that leads to $\expval{n_R(r)} = \expval{n_R(r+1)}$ reads
\begin{equation}\label{*0}
    F_{r+1} = \sum^r_{n=1} F_n/r.
\end{equation}
%
With $P_\text{det}^r = \sum^r_{n=1} F_n$, 
the transcendental equation Eq. (\ref{*0}) becomes 
\begin{equation}\label{*}
    P_\text{det}^r(\epsilon)  = r F_{r+1} (\epsilon),
\end{equation}
which yields a set of values of $\tau$ and $r$,
and as noted above this gives $\expval{n_R(r)} = \expval{n_R(r+1)}$.
Now exploiting the fact that the values of $r^*$ have a staircase structure, 
we start with $\epsilon=0$, and then $r^*=1$, 
increasing $\epsilon$, we get the transition point $r^*=1 \to r^*=2$, 
denoted by $1\to2$ transition.  
We then continue increasing $\epsilon$ to find the transition $2\to3$, etc. 
This means that the above formula yields the values of $r^*$ 
at sampling times given by $\epsilon$. 
This is a valid approximation in the case of large $\tau$ only as mentioned, 
and in this case, with Eq. (\ref{fna}) we have to solve for
\begin{equation}\label{**}
    \sum_{n=1}^r {1\over n} \cos^2 \left( 2 n \epsilon  - {\pi\over4} \right) 
    = {r \over r+1} \cos^2 \left[ 2 (r+1) \epsilon - {\pi\over4} \right], 
\end{equation}
and again we must increase $\epsilon$ from zero using $r=1$,  
to find the first transition $r^*=1 \to r^*=2$ at $\epsilon_{1\to2}$,  
then update to $r=2$ finding the transition point $\epsilon_{2\to3}$ etc. 
It is clear that when $r$ is very large, 
there is no solution to Eq. (\ref{**}). 
Since the left-hand side of Eq. (\ref{**}) 
can be simplified as $(1/2)\sum^{r}_{n=1} [1/n+ \sin(4n\epsilon)/n]$,
and in large $r$ limit, $\sum^{r}\sin(4n\epsilon)/n \to (\pi-4\epsilon)/2$,
and $\sum^{r}1/n \to \infty$ does not converge,
while the right-hand side is always bounded by $1$.
Physically it makes sense that we cannot witness the transitions forever,
since as mentioned $\expval{n_R(r)}$ is proportional to $r$ in large $r$ limit,
there must exist a minimum at finite $r$.
We cannot expect a restart strategy to be useful for $r\gg1$.
See Fig. \ref{ore0} for the comparison 
between exact results and approximations [Eq. (\ref{**}), Table \ref{trantau}].

\begin{figure}[ht]{}
\centering
\includegraphics[width=0.5\linewidth]{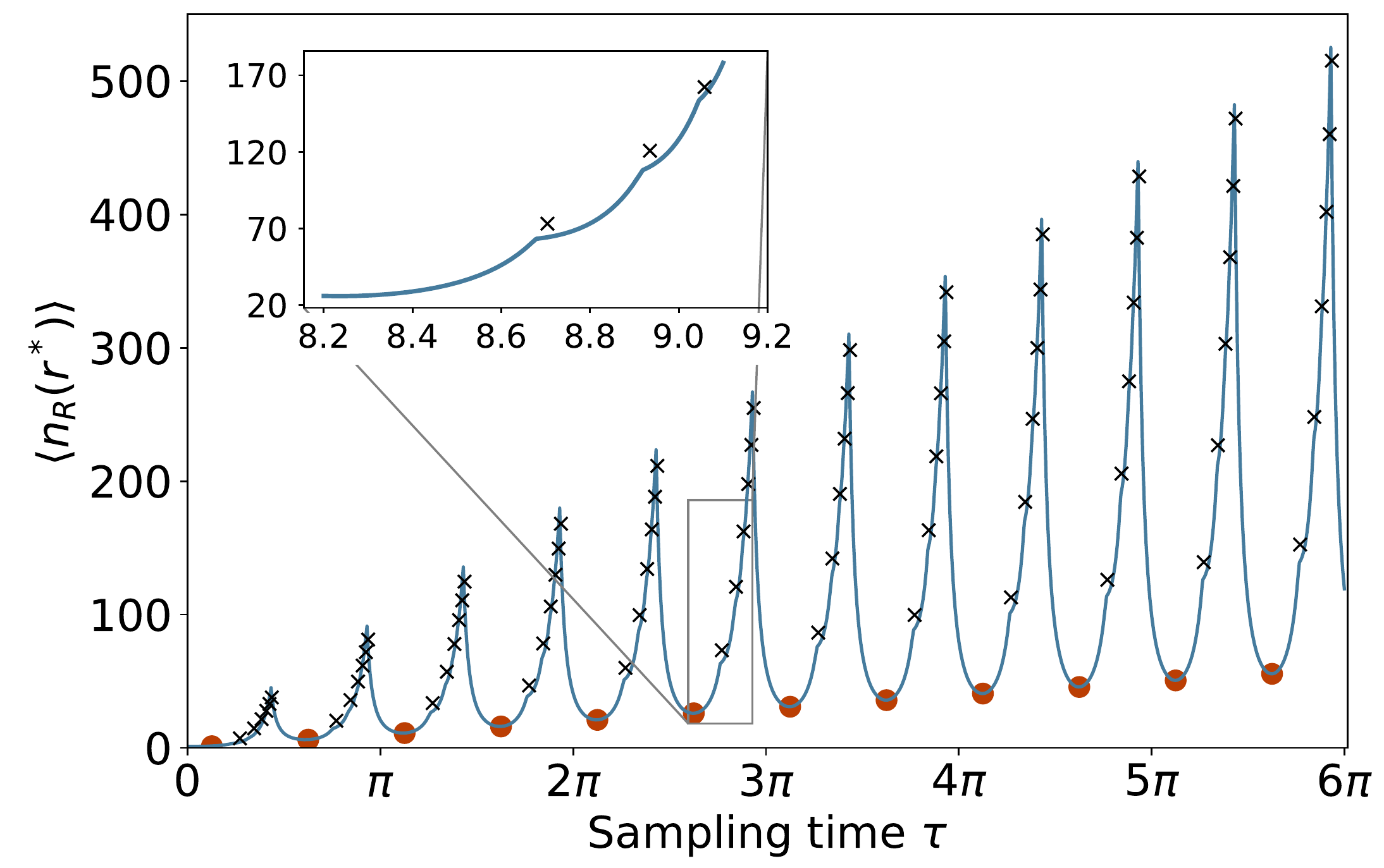}
\caption{The optimal mean $\expval{n_R(r^*)}$ vs. $\tau$ 
for $\delta=0$ (cyan curve).
The black crosses representing $\expval{n_R(r^*)}$ at transition $\tau$'s 
are plotted using Eq. (\ref{EQ3}). 
And at those transitions, nonsmoothness $\expval{n_R(r^*)}$ is witnessed (see the inset).
The minima of $\expval{n_R(r^*)}$ are predicted by Eq. (\ref{minima}) 
and presented by red closed circles.
As shown in the figure, 
$\expval{n_R(r^*)}$ exhibits large fluctuations 
and periodic-like behavior when the sampling time $\tau$ is varied. 
The peaks of $\expval{n_R(r^*)}$ are found close to the plunges of $r^*$ 
shown in Fig. \ref{Bess0}, 
namely close to the instability points 
like the one presented in Fig. \ref{tran}. 
See more details about the plot in Appendix \ref{appdix3}.
For an efficient search one clearly needs to consider the optimization 
of the process both with respect to the restart time, 
but also with respect to the sampling period $\tau$, and we will analyze this issue later.
}
\label{rmtend}
\end{figure}
%


After finding the optimal $r^*$, we study the mean $\expval{n_R(r^*)}$,
at the optimal choice $r^*$, 
which is of course the fastest way in mean sense to detect the particle. 
Note that when $F_n\ll 1$, 
using Eq. (\ref{eq13}),
we have 
\begin{equation}
    \expval{n_R(r^*)} \sim {r^*\over \sum^{r^*} F_n}.      
\end{equation}
From above arguments Eq. (\ref{*0}), at those transition $\tau$,
we have 
\begin{equation}
    \expval{n_R(r^*)} \sim {1 \over F_{r^*+1}}.
\end{equation}
This is obtained by calculating $F_n$, at the corresponding $\epsilon$,
namely, using Eq. (\ref{fna}),
\begin{equation}\label{EQ3}
    \expval{n_R(r^*)} \sim {(r^*+1)\pi\tau \over \cos^2 [2(r^*+1)\tau-\pi/4]}.
\end{equation}
See Fig. \ref{rmtend} for the numerical confirmation. 
The exact results are represented by the cyan line. 
And the theoretical $\expval{n_R(r^*)}$ at transition $\tau$'s 
calculated by Eq. (\ref{EQ3})
are represented by black crosses,
at which nonsmoothness of $\expval{n_R(r^*)}$ appears.
As $\tau$ is increased, 
the general trend is an increase of $\expval{n_R(r^*)}$, 
which is expected since the wave packet for large $\tau$ 
has spread out far from the detector when $\delta=0$. 
In addition to this trend we have a periodical set of maxima. 
Note the dramatic changes in those maxima presented in the figure. 
The maxima are at the plunge $\tau$'s, 
where $r^*$ falls from $6$ to $1$.
The minima of $\expval{n_R(r^*)}$
are actually the minima of $\expval{n_R(1)}=1/F_1$, 
when $r^*=1$,
namely,
\begin{equation}\label{eq24}
    \min(\expval{n_R(r^*)})=
    \min(\expval{n_R(r^*=1)}).
\end{equation}
%
Later we will discuss this again.
In large $\tau$ limit, using Eq. (\ref{fna}), 
we find when $\tau=\pi/8+k\pi/2$, 
$\max(F_1)=(\pi^2/8+k\pi^2/2)^{-1}$,
then the minima of $\expval{n_R(r^*)}$ is 
\begin{equation}\label{minima}
    \min(\expval{n_R(r^*)}) = {1\over \max(F_1)} 
    = {\pi^2 \over 8} + k{\pi^2\over2}.
\end{equation}
We plot in closed circles these theoretical minima 
indicated by Eq. (\ref{minima}).
Our theory Eqs. (\ref{EQ3},\ref{minima}) nicely matches the numerics, 
as shown in Fig. \ref{rmtend}.




\subsection{Nearest-neighbor detection: $\delta=1$}
Here we investigate the case 
where the detector is put at the neighboring site to the origin,
namely $\delta=1$.
In the Zeno regime, namely $\tau\to0$, using Eq. (\ref{meaneg}), 
we have 
\begin{equation}
    \expval{n_R(1)} \sim 1+\tau^{-2}, \;\;\; \expval{n_R(2)} \sim {5\over2}+\tau^{-2}, \;\;\;
    \expval{n_R(3)} \sim {14\over3}+\tau^{-2}, \cdots
\end{equation}
Hence $r^*=1$ when $\tau\to0$. 
The physical picture is the following:
for small $\tau$ we have a leakage of amplitude, from the starting point $x=0$, 
both to $x=1$ and to $x=-1$,
in fact the amplitudes at these states are the same. 
Now one tries to detect on $x=1$, and does not find the particle. 
One may choose to restart, 
which means that the amplitude at $x=-1$ restores to $x=0$. 
This benefits detection.
If looking at the asymptotics of the first-hitting probability,
with Eq. (\ref{eq1a}),
we find in the limit $\tau\to0$,
\begin{equation}
    F_1 \sim \tau^2-\tau^4, \;\;\; F_2\sim \tau^2-5\tau^4, \;\;\;
    F_3 \sim \tau^2-11\tau^4, \cdots
\end{equation}
So $F_1>F_2>F_3>\cdots$,
indicating that continuing with the measurement is not beneficial 
and a restart is the best option to speed up search. 
Hence performing restart after each failed measurement is the best strategy,
and $r^*=1$ when $\tau\to0$.

\begin{figure}[htbp]{}
\centering
\includegraphics[width=0.47\linewidth]{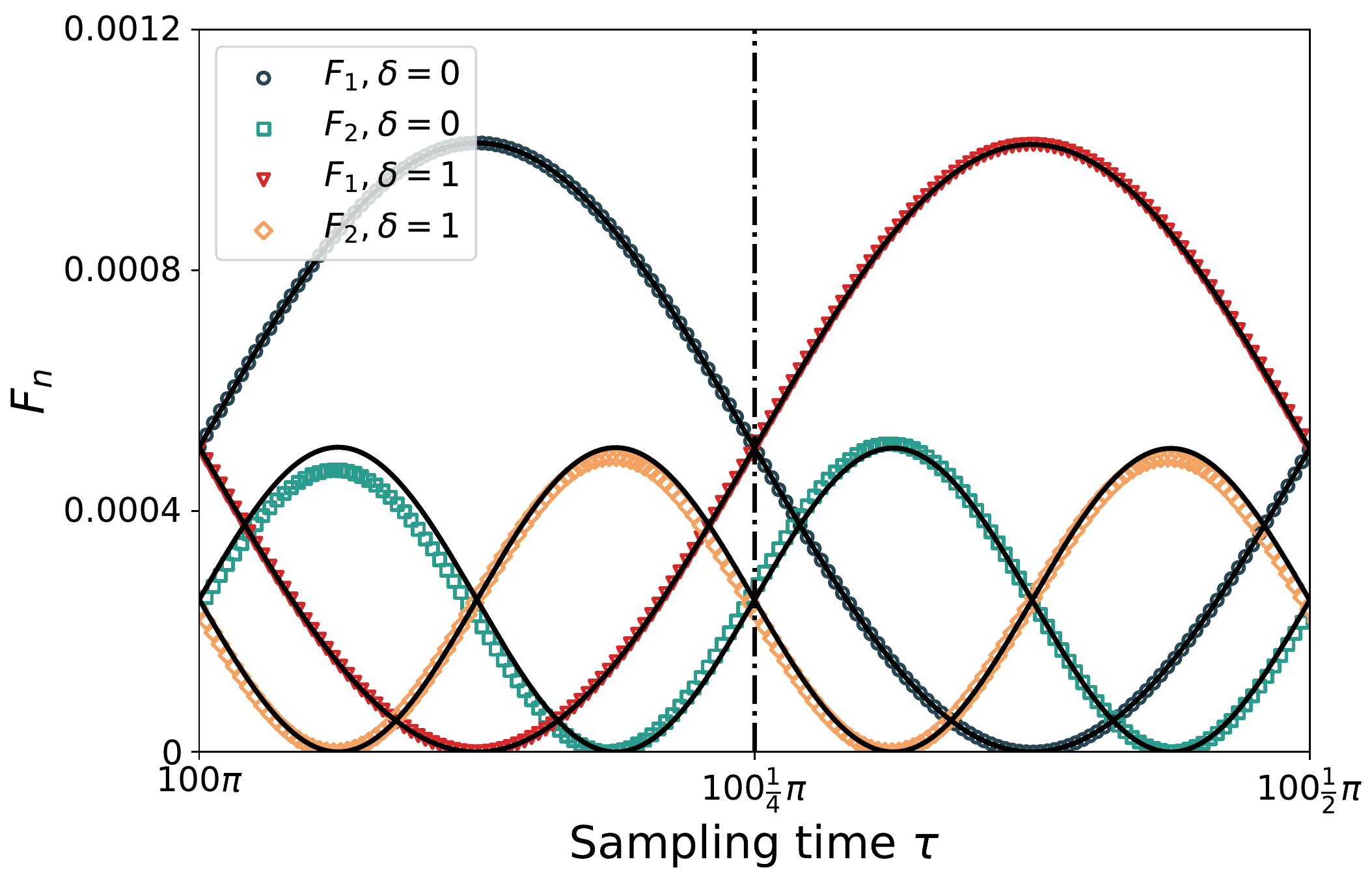}
\caption{$F_n$ vs. $\tau\in [100\pi, 100{1\over2}\pi]$ 
for $n=1,2$ and $\delta=0,1$.
The black curves represent the trigonometric approximations Eqs. (\ref{fna},\ref{fn1}).
As shown and mentioned in texts,
the $F_n$'s for $\delta=1$, in each interval $[\pi k/2, \pi(k+1)/2]$,
are symmetric to the ones for $\delta=0$ about $\pi k/2+\pi/4$ (the dashed line).
}
\label{fn12d12}
\end{figure}
\begin{figure}[ht]{}
\centering
\includegraphics[width=0.45\linewidth]{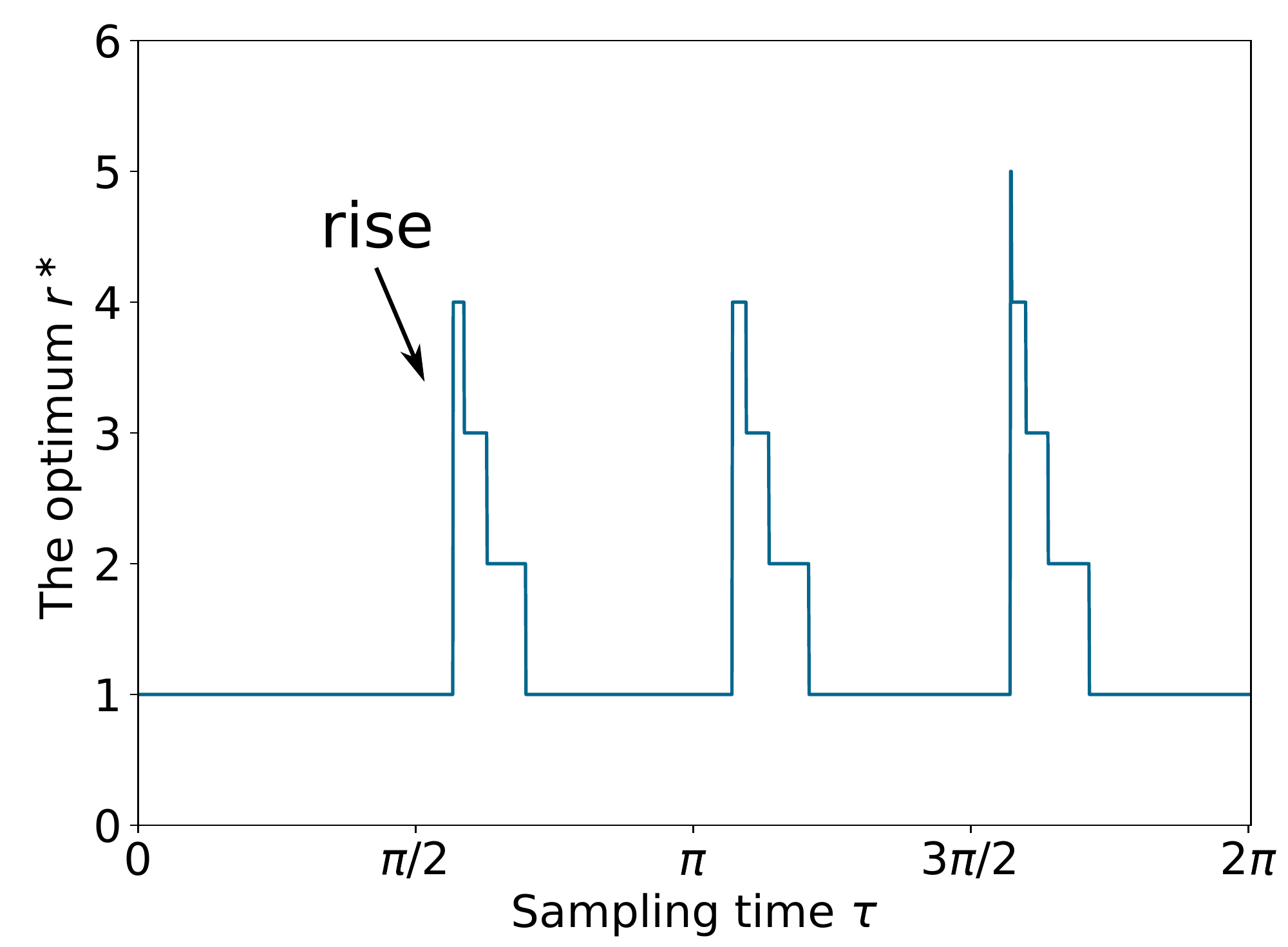}
\caption{The optimal restart time $r^*$ as a function of $\tau$ for $\delta=1$.
We see the staircase structure along with periodical rises, 
i.e. $r^*$ jumping from $1$ to $4$ or $5$ (see the arrow). 
The series of plateaus declines from the maximum from left to right 
unlike the case $\delta =0$ where we find the opposite trend [see Fig. \ref{Bess0}].
}
\label{or1qc}
\end{figure}
\begin{figure}[ht]{}
\centering
\includegraphics[width=0.45\linewidth]{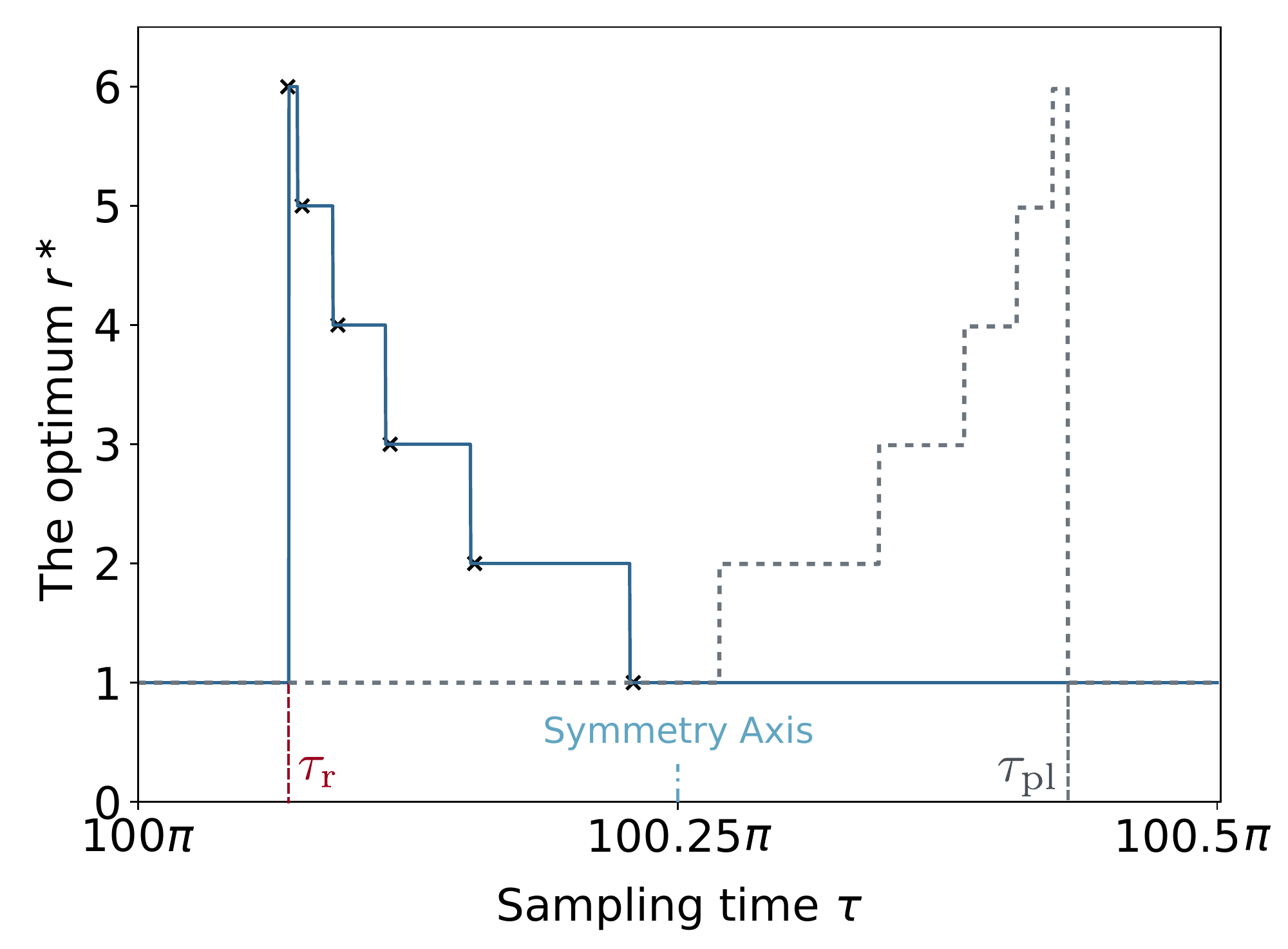}
\caption{$r^*$ versus $\tau$ with $\tau\in[100\pi, 100.5\pi]$ and $\delta=1$.
The cyan lines are the exact results obtained using Eq. (\ref{eq9}), 
and the black crosses represent the approximations 
[via applying $\epsilon\to\pi/2-\epsilon$ to Eq. (\ref{**})].
$\tau_\text{r}$ here is approximated as $k\pi/2+0.218$ with $k=200$. 
There is a mirror image of Fig. \ref{ore0} (the gray dashed line here), 
so the parity of the initial condition
determines the staircase structure which otherwise is universal 
in the sense that it does not depend on the initial condition. 
The symmetry axis for the two staircase patterns is $k\pi/2+\pi/4$ with $k=200$ here.
}
\label{orep1}
\end{figure}

%
In the opposite limit of large $\tau$, we have 
\begin{equation}\label{fn1}
    F_n \sim  {1\over n\pi\tau} 
              \cos^2 
              \left( 
                2n\tau-{3\pi\over4} 
              \right)
              .
\end{equation}
For $\tau=\pi k/2$, 
we find again that $F_n \sim (2n\pi\tau)^{-1}$,
and $r^*=1$, the same as Eqs. (\ref{fn2p}, \ref{eqadded}). 
To be more specific, since $F_n$ is a monotonic decaying function, 
for these special values of $\tau$ the best strategy is to restart 
after the first measurement. 
Similar to the case $\delta=0$,
we expect that increasing $\tau$ in every interval $[\pi k/2,\pi(k+1)/2]$
leads to the quanta jumps of $r^*$.
While it is noteworthy that if we let $\tau=\pi k/2-\epsilon$ with $k\gg1$,
Eq. (\ref{fn1}) becomes
\begin{equation}
    F_n \sim  {1\over n\pi\tau} 
              \cos^2 
              \left( 
                -2n\epsilon-{3\pi\over4} 
              \right)
        =     {1\over n\pi\tau} 
              \cos^2 
              \left( 
                2n\epsilon-{\pi\over4} 
              \right)
              ,
\end{equation}
which recovers to the case $\delta=0$.
This means that in the range $\tau\in[\pi k/2, \pi(k+1)/2]$, 
$F_n$ in the case of $\delta=1$ 
is symmetric to that in the case of $\delta=0$, 
with respect to $\tau=\pi k/2+\pi/4$, when $k$ is large,
as shown in Fig. \ref{fn12d12}.
Therefore,
we expect the behaviors of $r^*$ in the case $\delta=1$ 
is a mirror reflection to that in the case $\delta=0$ 
in every interval $[\pi k/2, \pi(k+1)/2]$,
with the symmetry axis at $\pi k/2+\pi/4$. 
Then the transition $\epsilon$'s are associated via 
$\epsilon \leftrightarrow \pi/2-\epsilon$ when $\delta$ switches between $0$ and $1$.
In Fig. \ref{or1qc} we plot $r^*(\tau)$ with $\tau\in[0,2\pi]$,
and the staircase structure is symmetric to the previous case, 
even when $\tau$ is not large.
In large $\tau$ limit, there appears definite symmetry (as seen in Fig. \ref{orep1}).

%

A calculation similar to that in Table \ref{ttau} 
is provided in Appendix \ref{tran1tau}.
And in the case $\delta=1$ we have a sudden rise in $r^*$, 
unlike the plunge for $\delta=0$. 
The rise $\tau$, denoted by $\tau_\text{r}$,
is equal to $\pi/2-\epsilon_\text{pl}\approx 0.218$.
And those transition $\epsilon$ are all symmetric 
to that in the case $\delta=0$ about $\pi/4$, as expected.
See Fig. \ref{orep1} for a comparison 
between the exact results and approximations.

Furthermore, one can readily show that in the case of even or odd $\delta$,
Eq. (\ref{fna}) or Eq. (\ref{fn1}) always hold, respectively,
and the pattern presented in Fig. \ref{orep1} is generic.
This insightfully indicates the $\delta$-independence of $F_n$ 
for all even/odd $\delta$, in large sampling time limit 
(similar properties are shown in Ref. \cite{Thiel2018a}). 
In other words, large $\tau$ renders $F_n$'s dependence 
only on the parity of distance between the initial and detected sites.
We will see this more clearly after the discussion on $\delta=2$.

\subsection{Next-nearest-neighbor detection: $\delta=2$} 
\begin{figure}[ht]{}
\centering
\includegraphics[width=0.45\linewidth]{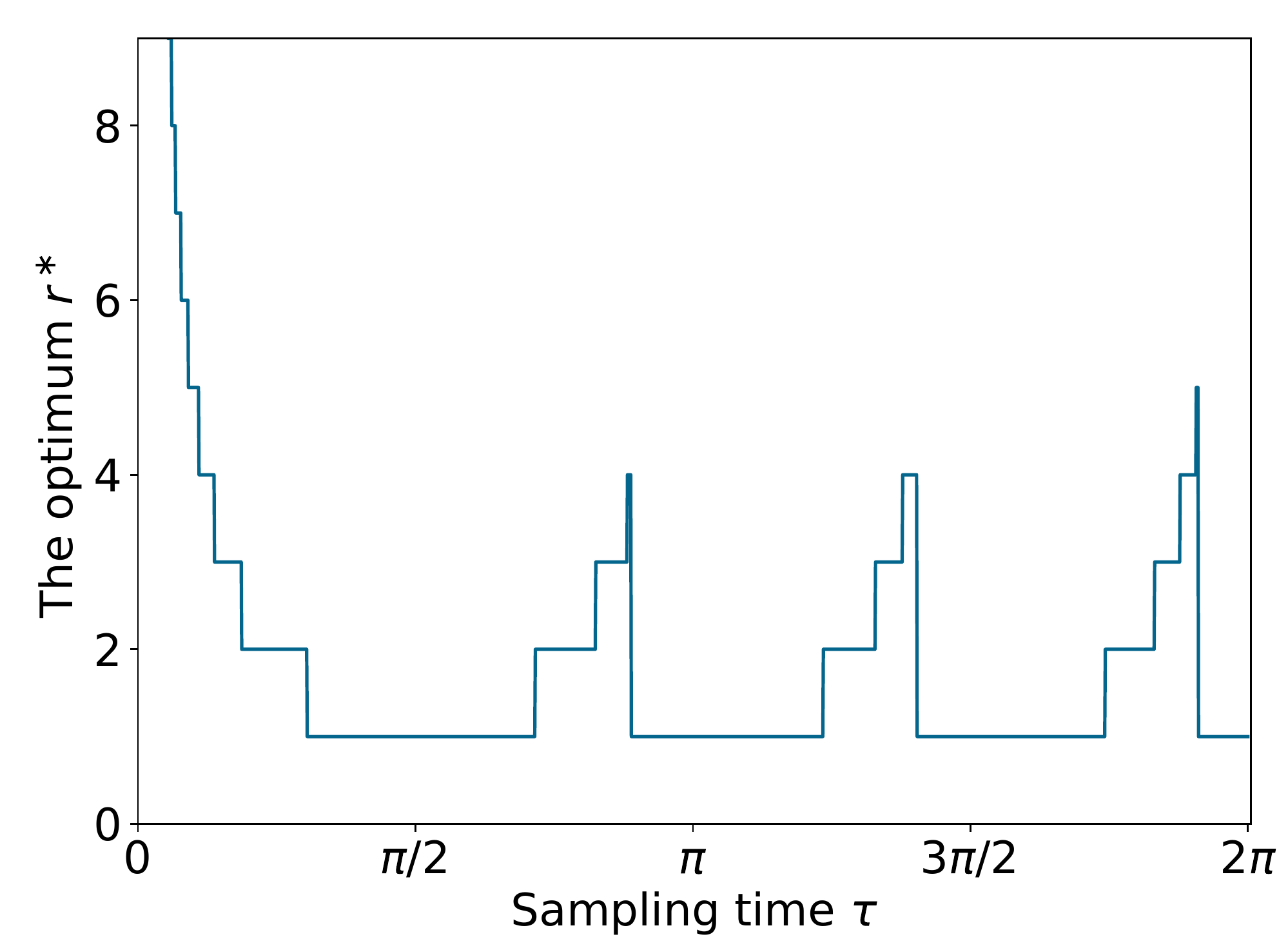}
\caption{The optimal restart time $r^*$ as a function of $\tau$ for $\delta=2$.
We see the staircase structure along with periodical plunges. 
}
\label{or2qc}
\end{figure}

When $\delta=2$, in the limit of $\tau\to0$,
we find using Eq. (\ref{eq1a})
\begin{equation}
    F_1 \sim {1\over4}\tau^4-{1\over6}\tau^6, \;\;\; F_2\sim {9\over4}\tau^4-6\tau^6, \;\;\;
    F_3 \sim {25\over4}\tau^4 - {235\over 6}\tau^6, \cdots
\end{equation}
Namely $F_1<F_2<F_3<\cdots$ till some $n\sim1/\tau$ and then decreases. 
This behavior is very different if compared with the cases 
$\delta=0$ and $\delta=1$, 
where $F_1$ was the maximum of the set $\{F_1, F_2, F_3, ...\}$.
We expect that $r^*$ exhibits divergence in this limit, 
see Fig. \ref{or2qc}.
Checking the asymptotics of $\expval{n_R(r)}$,
with Eq. (\ref{meaneg}), 
we have
\begin{equation}
    \expval{n_R(1)} \sim 4\tau^{-4}, \;\;\; \expval{n_R(2)} \sim {4\over5}\tau^{-4}, \;\;\;
    \expval{n_R(3)} \sim {12\over35}\tau^{-4}, \cdots
\end{equation}
Hence $\expval{n_R(1)} > \expval{n_R(2)} > \expval{n_R(3)} > \cdots$.
We use the theory in Ref. \cite{Ruoyu2023}, to study the Zeno limit.

\begin{figure}[htbp]{}
\centering
\includegraphics[width=0.45\linewidth]{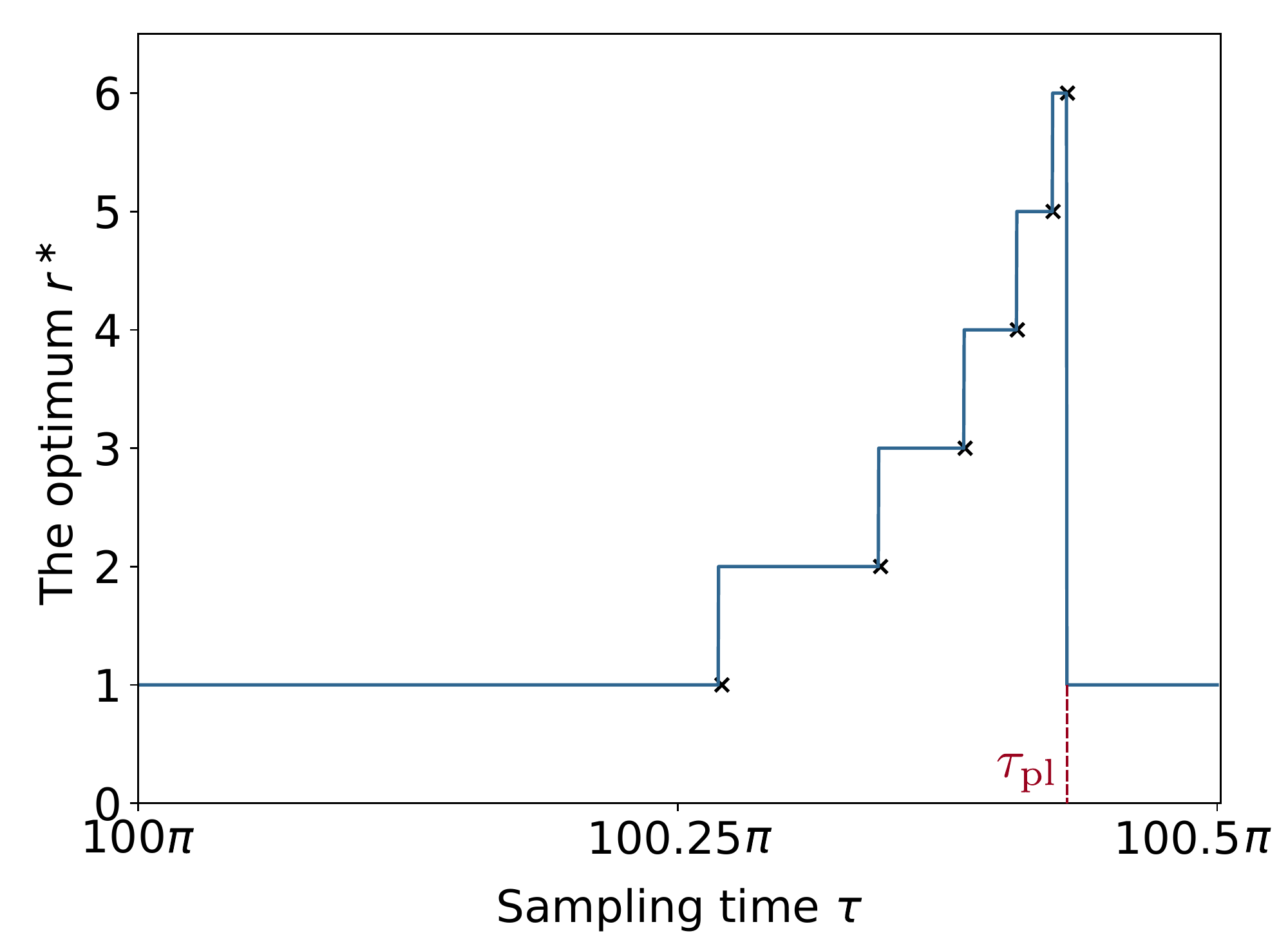}
\caption{$r^*$ versus $\tau$ with $\tau\in[100\pi, 100.5\pi]$ and $\delta=2$.
The cyan lines are the exact results, 
and the black crosses represent the approximations using Table \ref{trantau}.
$\tau_\text{pl}$ here is approximated as $100\pi+1.353$. 
}
\label{orep2}
\end{figure}

Assuming large $\tau$, $F_n$ is re-expressed as \cite{Friedman2017a,Thiel2018a}
\begin{equation}
    F_n \sim  {1\over n\pi\tau} 
              \cos^2 
              \left( 
                2n\tau-{\pi\over4} 
              \right)
              .
\end{equation}
This is the same as in the case $\delta=0$,
since the initial condition $\delta$ 
only affects the phase in $F_n$ via $\pi \delta/2$,
and then the same parity of $\delta$ leads to the same $F_n$,
and to the same pattern of $r^*(\tau)$.
Thus we could use Table \ref{trantau} to approximate $r^*$'s staircase structure in this case, see Fig. \ref{orep2}.

\subsection{The parity of $\delta$ matters}
\label{parityM}
Following above discussion, we note a remarkable feature of 
the quantum first-hitting times probabilities, 
namely that beyond a phase, they are independent of $\delta$:
\begin{equation}\label{eq29}
\begin{aligned}
    \text{For Even $\delta$:}\quad 
    F_n &\sim  {1\over n\pi\tau} 
                \cos^2 
                \left( 
                2n\tau-{\pi\over4} 
                \right)
                , \\
    \text{For Odd $\delta$:}\quad
    F_n &\sim  {1\over n\pi\tau} 
                \cos^2 
                \left( 
                2n\tau-{3\pi\over4} 
                \right)
                .
\end{aligned}
\end{equation}
This is obtained with the large argument asymptotics for $J_n(x)$,
which was also used in the case of return in Ref. \cite{Friedman2017a},
and in the supplemental material of Ref. \cite{Thiel2018a}.  
Eq. (\ref{eq29}) is valid in the large $\tau$ limit,
and we note that when $n$ is large, 
the first-detection probability $F_n$ transits to a $n^{-3}$ power-law decay
\cite{Friedman2017a,Thiel2018a},
but this does not affect our theory, which only focuses on finite $r$ restart.

Hence, based on Eq. (\ref{eq29}), the staircase patterns, 
are {\em binary} and merely determined by the parity of $\delta$.
These two patterns are mirror reflection to each other,
connected by the operation/mapping 
$\epsilon_{k+1\to k} \leftrightarrow \pi/2-\epsilon_{k\to k+1}$.
The origin of universality of staircase, is related to the fact 
that for large $\tau$, the first-detection amplitude is directly 
associated to the wave function of the measurement-free process. 
This then leads to specific phases in the asymptotic expansion Eq. (\ref{eq29}) 
which depends on the parity only.
We note that related work on discrete-time quantum walks also found interesting effects for the parity
on the probability distribution of a walker's position, see Refs. \cite{Xu2010,Omanakuttan2021}. 

To put it differently,
a unity change of the distance between the target and source $\delta$, in units of the lattice constant,
results in a ``flipping'' of the staircase pattern of the optimal restart time,
which again indicates, in our view, a type of instability.
This effect was demonstrated in Figs. \ref{ore0},\ref{orep1},\ref{orep2}. 

\section{The robustness of instability against perturbations in $\tau$}
\label{robust}
%
In this section, we discuss the effects of precision of the sampling time $\tau$
on the instability.
In experiments, 
one cannot perfectly achieve the stroboscopic measurement protocol, 
namely, there could be some random variation around the preset sampling time or measurement period $\tau$.
Hence, the practical sampling time, denoted by $\tilde{\tau}$, is fluctuating around $\tau$.
This randomness in the ideal periodicity of measurements,
would have non-negligible influences, when becoming considerable. 
In particular such noise could possibly modify the probability $F_{n}$'s oscillatory behaviors 
\cite{kessler2021},
and thereby probably eliminates the instability exposed in this paper. 
Will the instability be present when the noise is weak? 
And how much randomness will eliminate the instability?
These issues are addressed below using numerical methods.

Without loss of generality, 
we chose uniformly distributed deviation from the ideal sampling time $\tau$.
Specifically, the relative deviation $|\tilde{\tau}-\tau|/\tau$ 
is uniformly distributed in the interval $[-w, w]$, 
with $w$ chosen as $0.1$, $0.2$, and $0.3$ in this work.
Namely, the actual $\tilde{\tau}$ is uniformly distributed 
within $[\tau(1-w), \tau(1+w)]$,
and hence $w$ characterizes the noise level affecting the precision of sampling time.  
Inspired by the unstable behaviors in the vicinity of the plunge $\tau$,
where we witnessed two minima competing with each other, 
as demonstrated in Fig. \ref{tran},
we chose the ideal $\tau=1.35$ and the return case $\delta=0$,
and observe how the noise level $w$ affects the behaviors of those two minima.

It is noteworthy that 
we numerically study the issue 
for quantum hitting times under restart with random $\tilde{\tau}$
and Eq. (\ref{eq9}) is invalid 
since the denominator $P^r_\text{det}=\sum_{n=1}^{r}F_{n}$ is not constant,
leading to the number of restart not obeying a geometric distribution anymore 
\cite{Ruoyu2023}.
Thus, we employ the Monte-Carlo method to perform simulations. 
The procedures are described as follows:

%
%
\begin{enumerate}[(i)]
    \item   {\em Initialization of the quantum walker:} 
            The quantum walker is initially 
            evolved from a predefined state in accordance with 
            the Schrödinger equation. 
            This evolution occurs over a time duration, $\tilde{\tau}_1$, 
            which is a uniformly random variable within the range 
            $[\tau(1-w), \tau(1+w)]$.
    \item   {\em Random coin tossing for detection assessment:} 
            A random variable, referred to as a ``coin'', is generated. 
            This variable is uniformly distributed within the interval $[0, 1]$. 
            The purpose of the coin is to ascertain whether the quantum walker 
            is detected following the initial state's evolution. 
            This determination is made by comparing the coin's value 
            with the detection probability, 
            which is derived from the unitary evolution.
    \item   {\em Non-detection and state modification:} 
            If the coin value falls below the computed detection probability,
            we are done and the hitting time is $1$.
            If the coin value exceeds the computed detection probability, 
            it signifies that the walker remains undetected. 
            In this case, the amplitude at the target site $\ket{0}$ is erased, 
            and the wave vector is renormalized.
            Subsequently, the single-site-erased wave vector undergoes unitary evolution for a duration, $\tilde{\tau}_2$. 
            Notably, $\tilde{\tau}_2$ is an independent and identically distributed 
            (i.i.d.) random value, akin to $\tilde{\tau}_1$. 
            The objective is to compute the probability of detection 
            at the time $t = \tilde{\tau}_1 + \tilde{\tau}_2$.
    \item   {\em Repeated detection attempts: }
            Post the initial non-detection, a second i.i.d. coin is generated 
            and compared with the newly computed detection probability 
            to decide if the walker is detected at this stage, 
            as in the step (iii).
    \item   {\em Criteria for repetition termination under sharp restart:} 
            The process iterates until the coin value is less than 
            the computed probability of detection, 
            marking the end of a repetition cycle. 
            Alternatively, if the process extends up to 
            a preset fixed restart step, $r$ 
            (i.e. after a cumulative time of $t=\tilde{\tau}_{1}+\tilde{\tau}_2+\cdots+\tilde{\tau}_{r}$),
            and the walker remains undetected, 
            the entire procedure recommences from the initial state, 
            repeating the procedures (i)-(v).
    \item   {\em Restarted hitting time calculation: }
            Once we get, for the first time, the scenario 
            where the coin value falls below the computed probability of detection
            at the corresponding time, the process is done.
            The number of all preceding unsuccessful attempts, incremented by $1$,
            is recorded as the first-detection time, or the hitting time, 
            under the restart condition
            \footnote{
            Our measure of first-detection is based on the number of measurements made, which for $w=0$ is the measurement time divided by the fixed $\tau$ 
            (neglecting fluctuations). 
            For $w\neq 0$ one can also record the time till the first-detection, namely the sum of all the waiting times, until the first-detection
            }.
    \item   {\em Realizations and expected value determination: }
            The aforementioned procedures, executed for obtaining 
            a single value of the hitting time $n_R$ under $r$-step restarts, 
            is called a single realization. 
            To ascertain the expected value of $n_R$ as a function of $r$,
            large number of realizations are conducted for each value of $r$.
\end{enumerate}
\begin{figure}[htbp]{}
\centering
\includegraphics[width=0.55\linewidth]{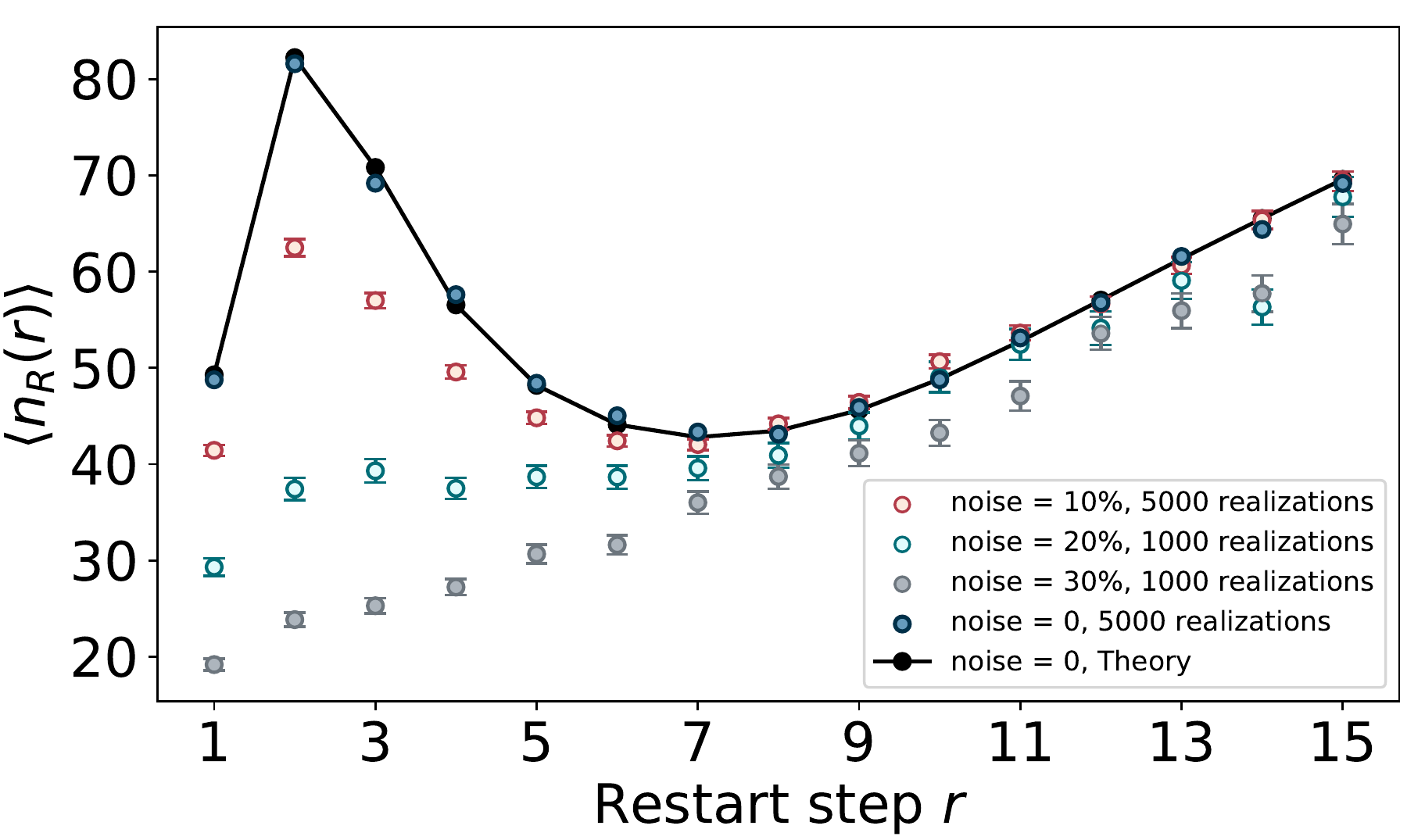}
\caption{
The mean hitting time under restart 
$\expval{n_R(r)}$ versus 
restart step $r$ for 
different levels of noise characterized by the width of the distribution of $\tilde{\tau}$.
We see the appearance of two minima of $\expval{n_R(r)}$ 
when the noise is chosen as $0.1$, $0.2$ fluctuating around $\tau=1.35$, 
while the absence of two minima when the noise is $0.3$.
Hence the instability is quite robust under noise in the sampling time.
The zero-noise data is for confirmation 
of the validity of the theory Eq. (\ref{eq9}).
The simulations are obtained using Monte-Carlo methods as described in the text.
}
\label{fig14}
\end{figure}
%
This was implemented with a {\em Python} program, 
generating the results presented in Fig. \ref{fig14},
for three different levels of noise $w$, i.e. 
$0.1$, $0.2$, $0.3$.
We note that, since computers cannot simulate the dynamics of an infinite line,
practically, a line of $120$ sites was used to approximate the unbounded model, 
and the maximal restart step is chosen as $15$ to ensure 
that the boundary does not affect the dynamics of the wave packet 
prior to restart (i.e. $15\tau(1+0.3)\cdot\max(v_g)=15\times1.35\times1.3\times2=52.7<120/2$, where as mentioned the group velocity $v_g=\partial_k E(k)$, and the initial site is set in the middle of the line).
As seen, the instability is robust,
in the sense that 
the presence of two minima is clearly visible for $w=0.1$ and $w=0.2$, indicating that the basic phenomenon 
is immune to noise. 
For $w=0.3$ the existence of a pair of minima is somewhat vague, 
and hence roughly when $w=0.3$, the effect we have found in the main text is wiped out.

\section{Optimization with respect to both $\tau$ and $r$}
\label{DoubleOpt}
In this section we will not deal with the noise problem.
All the above discussions focus on the optimal restart time for given sampling time $\tau$,
and a natural question arises from the existence of a globally optimal choice 
$(\tau^\ast,r^\ast)$ that achieves a global optimization of the mean $\expval{n_R}$,  
with regard to both controlling parameters.
We note here that the number of measurements till the first-detection is minimized in this work, and the global optimization of the expected time is left for future study.

\begin{figure}[htbp]{}
\centering
\includegraphics[width=0.49\linewidth]{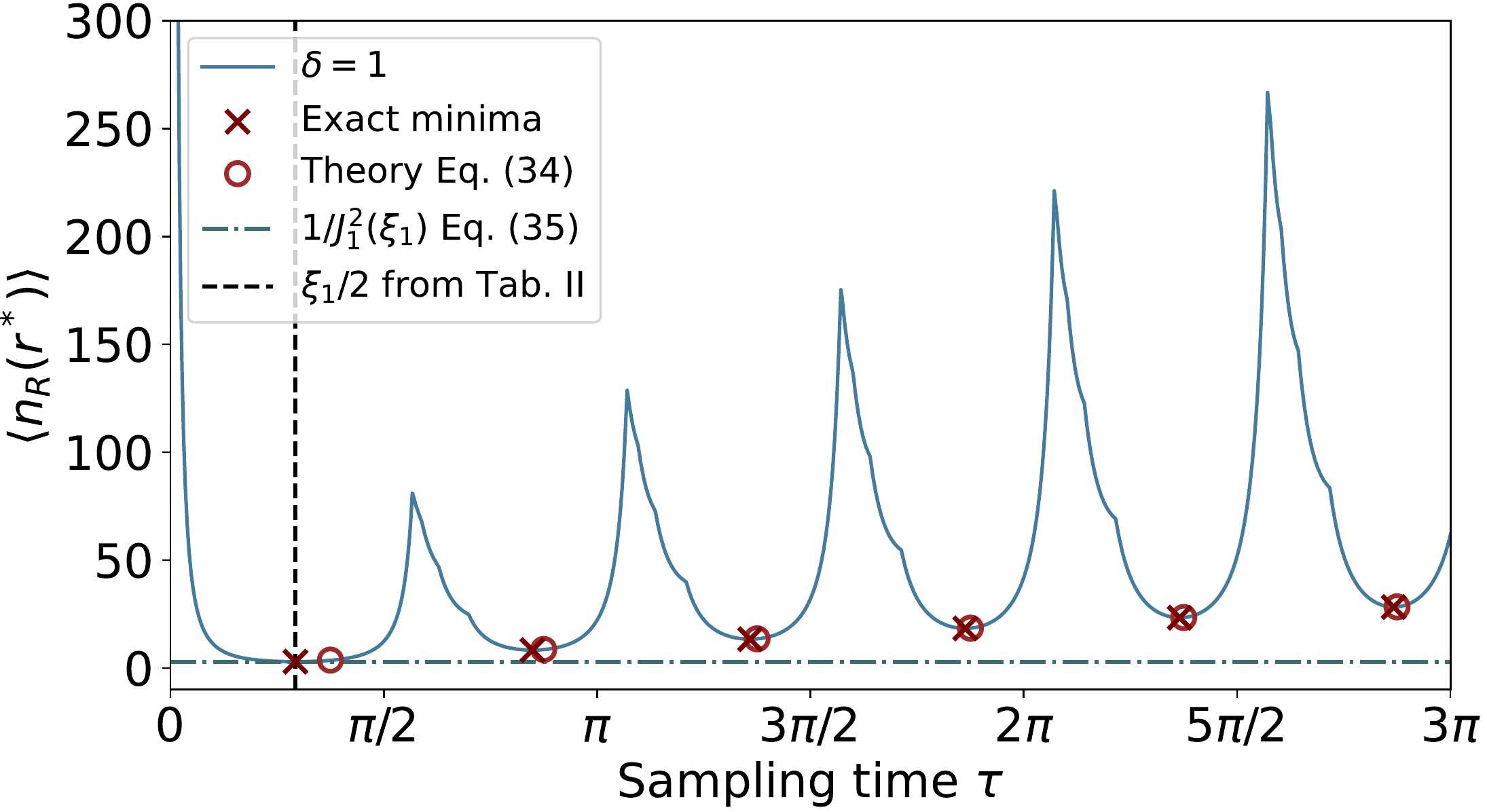}
\includegraphics[width=0.49\columnwidth]{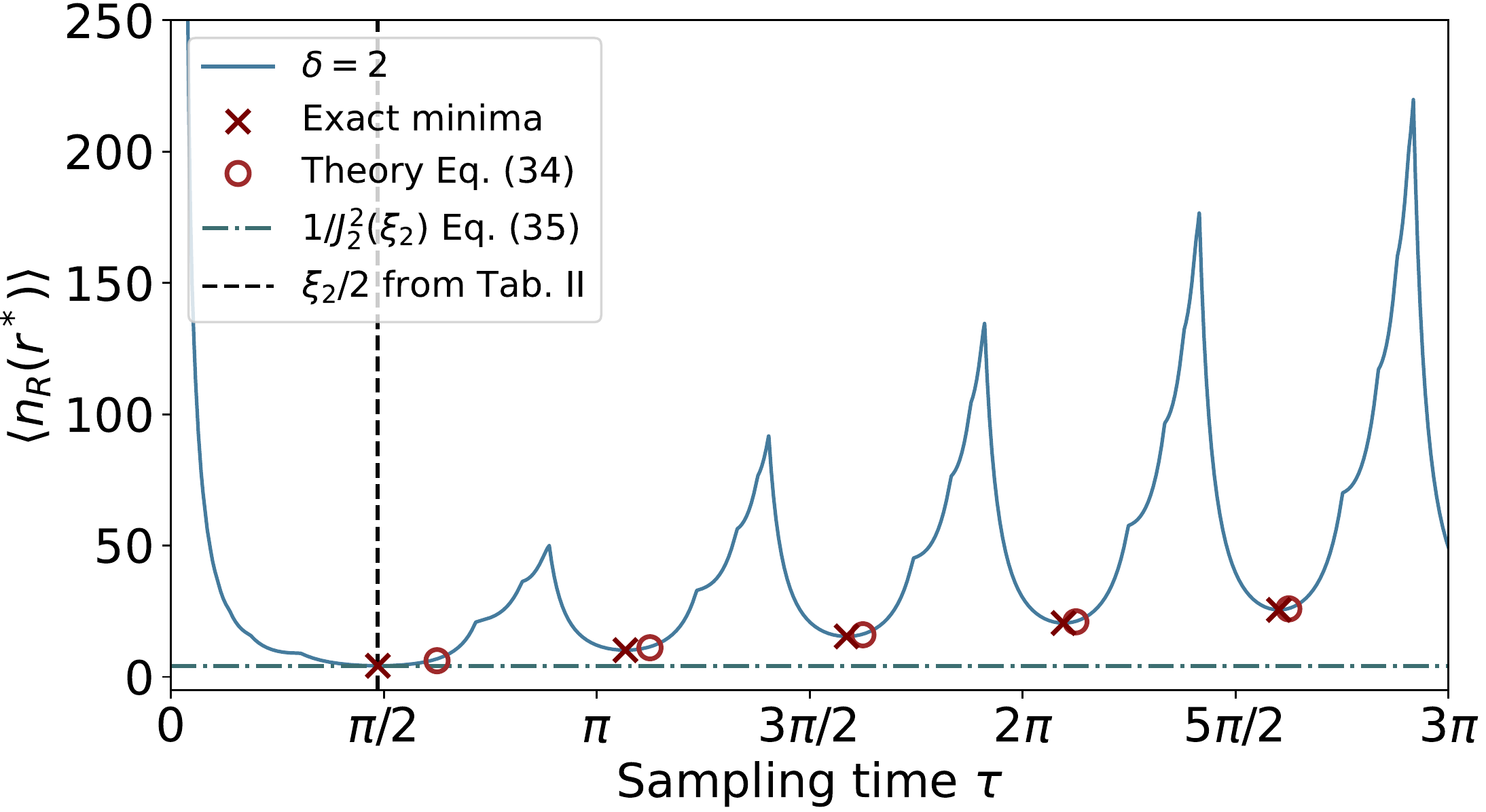}
\caption{
The optimal mean $\expval{n_R(r^*)}$ vs. $\tau$ 
for $\delta=1$ (left) and $\delta=2$ (right).
%
The numerically-obtained 
exact minima (crosses) 
converge to the theoretical results (open circles) calculated with Eq. (\ref{minimadelta})
as $\tau$ becomes larger.
We see that the global optimization (leftmost cross) with respect to 
both $\tau$ and $r$ 
is achieved for $\tau=\xi_\delta/2$ 
(vertical dashed lines, 
obtained from Eq. (\ref{eq33}) and Tab. \ref{tintd}) 
that minimizes $\expval{n_R(1)}
=[J_\delta(2\tau)]^{-2}$ 
(horizontal dashdotted lines),
i.e. 
when $F_1(\tau)=J^2_\delta(2\tau)$ reaches its global maximum.
The result clearly shows that 
Eq. (\ref{eq33}) is an excellent approximation.
}
\label{minrm}
\end{figure}

The function $\expval{n_R(r^*)}$ has multiple minima, 
as shown already in Fig. \ref{rmtend}.
We also present $\expval{n_R(r^*)}$ for cases $\delta=1$ and $\delta=2$
in Fig. \ref{minrm}.
%
As mentioned in Eq. (\ref{eq24}) and likewise here, 
in the large $\tau$ limit which validates the $J_\alpha(x)$ approximated as trigonometric functions,
using Eq. (\ref{eq29}),
all the minima of $\expval{n_R(r^*)}$ are
found at $\tau=\tau^\dag$ minimizing 
$\expval{n_R(r^*=1)}$, i.e. 
\begin{equation}\label{minimadelta}
    \text{min}(\expval{n_R(r^*)}) = 
    \pi \tau^\dag,
    \text{ with } \tau^\dag = (-1)^\delta \pi/8 + k\pi/2,
    \text{  and  }
    r^* = 1.
\end{equation}
Here $k$ is a large integer.
Eq.~(\ref{minimadelta}) is valid for large $k$ or $\tau$. 
It is not surprising that 
in this large $\tau$ limit 
$r^*=1$, 
since the wave packet has a long time to evolve,  
and hence statistically 
the particle is far from 
the target after and before 
the first measurement, 
thus it is wise to restart.
Unfortunately, or not, 
the global minimum as found, 
for example in Fig.~\ref{minrm}, 
is found for small $\tau$,
provided that $\delta$ is also small. 
For large $\delta$, 
as we discuss below, special features emerge.

%
\begin{figure}[htbp]{}
\centering
\includegraphics[width=0.49\linewidth]{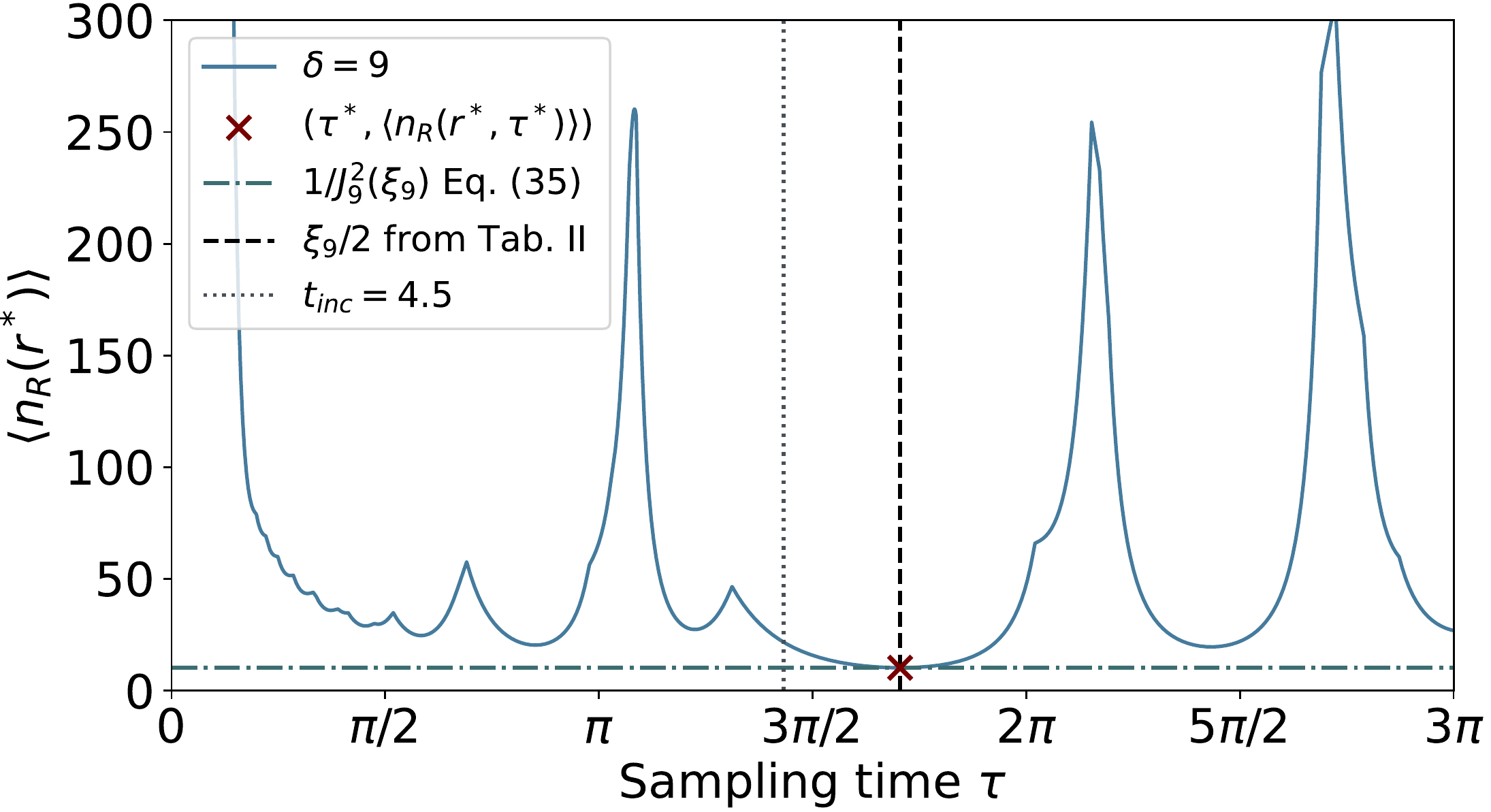}
\includegraphics[width=0.49\linewidth]{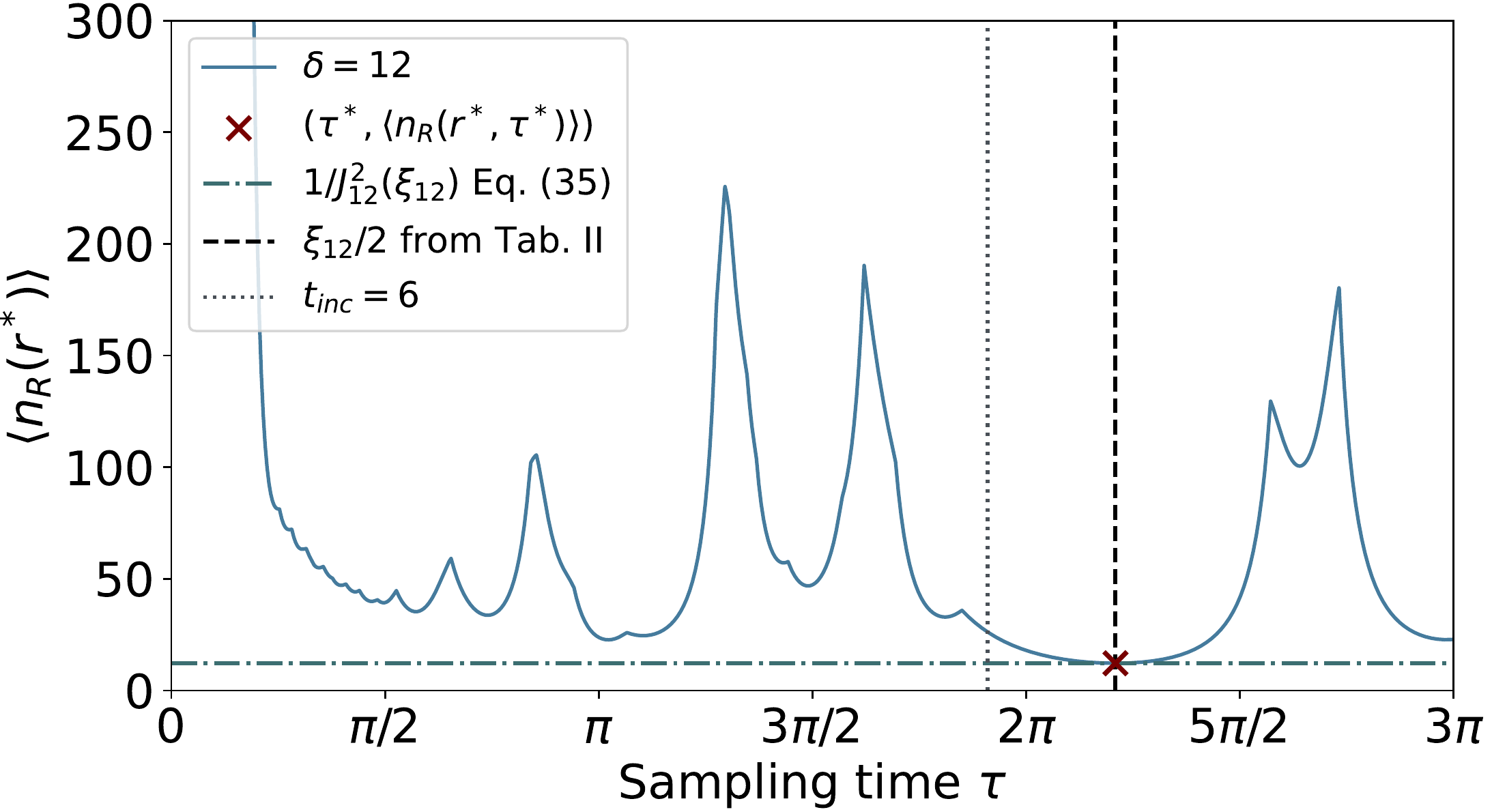}
\caption{
The optimal mean $\expval{n_R(r^*)}$ vs. $\tau$ for $\delta=9$ (left) and $\delta=12$ (right).
We see that the numerical global minimum of $\expval{n_R(r^*)}$ (crosses) is matched by our theory Eq. (\ref{eq33}), namely $\expval{n_R(1,\xi_\delta/2)}
=[J_\delta(\xi_\delta)]^{-2}$ 
(see the vertical dashed lines and horizontal dashdotted lines, obtained from Tab. \ref{tintd}). 
The incidence time $t_\text{inc}=\delta/2$ (vertical dotted lines)
also provides a rough approximation for $\tau^*$.
Thus, we conjecture that the most efficient detection strategy, 
in the sense of the least number of measurement attempts,
is to restart after the first measurement with the largest success probability in the first measurement.
}
\label{minrm2}
\end{figure}
%

\begin{figure}[htbp]{}
\centering
\includegraphics[width=0.6\linewidth]{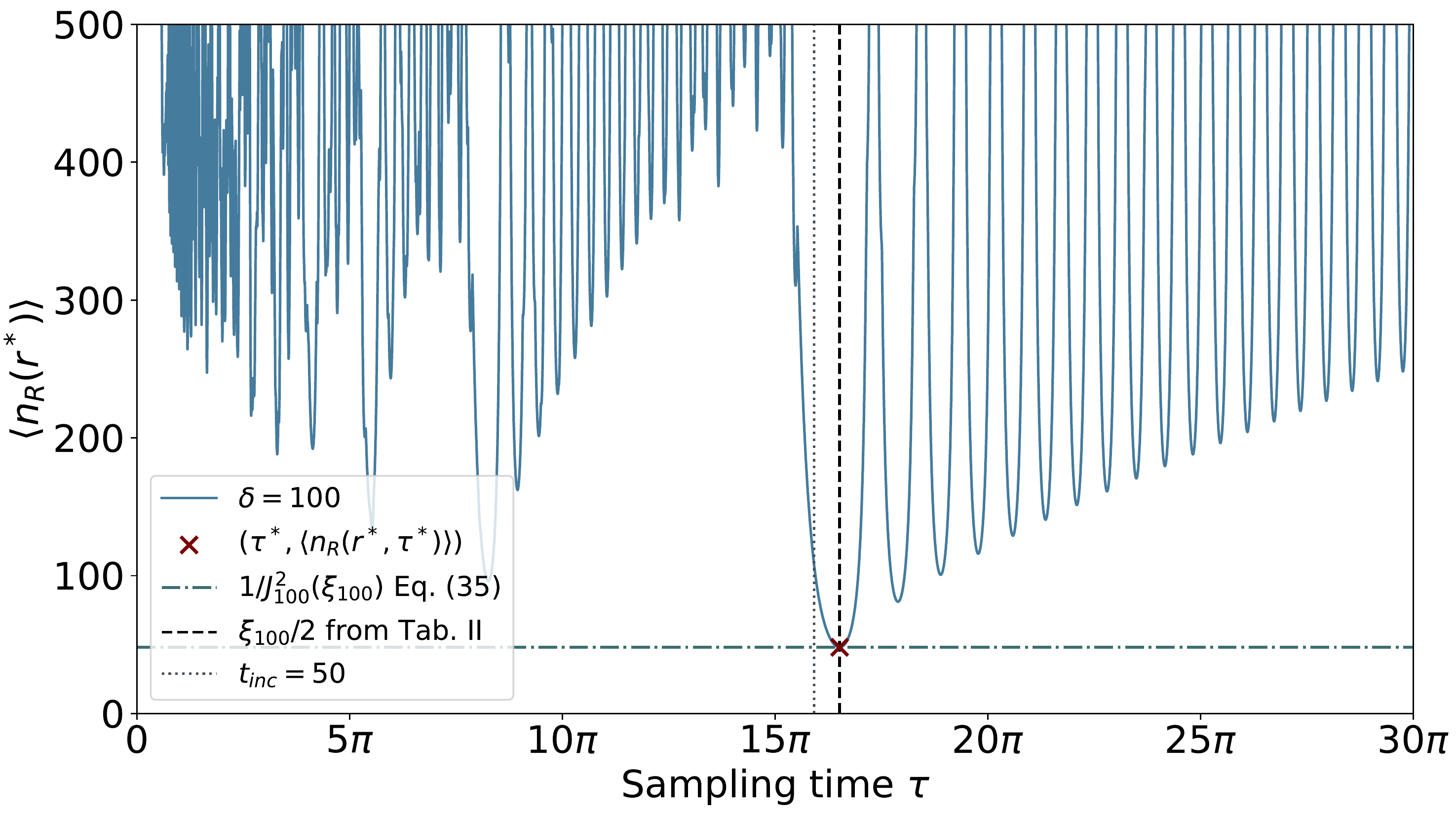}
\caption{
The optimal mean $\expval{n_R(r^*)}$ vs. $\tau$ for $\delta=100$.
We see that the exact global minimum of $\expval{n_R(r^*)}$ (cross) 
is matched by the minimum of $\expval{n_R(1)} 
= [J_{100}(2\tau)]^{-2}$ 
(the horizontal line) at $\tau=\xi_{100}/2$ 
(the vertical dashed line, 
see Tab. \ref{tintd}), 
and the incidence time $t_\text{inc}=\delta/2$ (vertical dotted line) gives pretty good approximation for $\tau^*$.
}
\label{minrm100}
\end{figure}

%
We conjecture, 
that even beyond the large $\tau$ limit, the global minimum is 
found for $r^*=1$. 
More specifically, 
let $r^*$, $\tau^*$ 
be the sampling time 
and restart time 
that minimize $\expval{n_R}$, 
and we suggest:
\begin{equation}\label{eq33}
    \expval{n_R(r^*,\tau^*)} = {1\over F_1(\tau^*)} = {1\over \max(F_1)}.
\end{equation}
Recall that $F_1=J^2_\delta(2\tau)$ (see Eq. (\ref{eq1a})),
thus, $r^*=1$ and $\tau^*=\xi_{\delta}/2$,
where $t=\xi_{\alpha}$ marks the highest peak of $J_\alpha(t)$,
give the global minimum of 
$\expval{n_R}$.
Eq. (\ref{eq33}) should hold for any value of $\delta$. 
Nevertheless, an estimate for $\tau^*$, which becomes more accurate for large $\delta$, 
can be found in the recently published literature.
Refs. \cite{Mecholsky2021,DLMF2021} show that $2\tau^*=\xi_{\delta}\sim \delta$,
whose numerical confirmation, 
as well as exact numerical solutions for other extrema of $J_\alpha(t)$
(also used in Fig. \ref{minrm}, the data presented by crosses),
can be found in the supplementary material of Ref. \cite{Mecholsky2021}.
Thus, the global minimum is given by 
\begin{equation}\label{globalOpt}
    r^*=1, \quad \tau^*\sim {\delta\over2} 
    =
    {\delta \over \max(v_g)} =:t_\text{inc},
\end{equation}
where $t_\text{inc}=\delta/2$ is the incidence time (in time units with $\gamma=1$) 
during which the wave-front travels 
from $\ket{0}$ to $\ket{\delta}$ (see Fig. \ref{fnn}). 
Hence, as mentioned, 
this suggests that 
$\tau^* \simeq \delta/2$ 
becomes a better approximation as $\delta$ grows,
namely for large $\delta$, 
$t_\text{inc}$ gives a good estimate of the globally optimized sampling time $\tau^*$.
This
again manifests the ballistic spreading of the wave packet's wave-front.
See Fig. \ref{minrm2} where we witness Eq. (\ref{globalOpt}) also works well for not too large $\delta$, i.e. $\delta=9,12$. 
As expected, for $\delta=100$, presented in Fig. \ref{minrm100}, the approximation works even better,
see also Tab. \ref{tintd}.
The global optimum is physically interpretable, 
since this special $\tau$ allows the largest part of the wave packet to arrive at the target state, 
once collapsed, it is best to start anew, namely a restart.
For small $\delta$ say $1,2$, 
Eq. (\ref{eq33}) is tested 
in Fig. \ref{minrm}. 
For this test 
we find the maximum of $|J_\delta(2\tau)|$ semianalytically with a simple program.
We see that 
Eq. (\ref{eq33}) is valid for small 
$\delta$, 
and while it holds also for large $\delta$, 
Eq. (\ref{globalOpt}) is simpler.
We will now analyze the reliability of our conjecture Eq. (\ref{eq33}), 
where the key is whether the global minimum always occurs for $r=1$.

%

When $\delta$ is large, 
namely it takes a long time to reach the target, 
we expect that large $\tau$ is useful.
We can prove that in the large $\tau$ limit, 
the minima of $\expval{n_R(r^*)}$ is always when $r^*=1$.
To do this,
one needs to justify that the minima of $\expval{n_R(1)}$ are smaller than 
those of $\expval{n_R(r)}$ for any $r\ge2$, in every interval of resemblance.
Recall that for large $\tau$, 
the mean detection time exhibits periodic-like behaviors, 
see Fig. \ref{rmtend},
and hence we have intervals or resemblance.
With Eq. (\ref{meaneg}),
the equivalent statement is that the maximum of $J_\delta^2(2\tau)$
is larger than that of $1/\expval{n_R(r)}$ for any $r\ge2$,
hence in the large $\tau$ limit approximation
\begin{equation}\label{mlm}
    \max \left[ {1\over \expval{n_R(1)}} \right] 
    = \max \left[ J_\delta^2(2\tau) \right] 
    > \max \left[ {\sum_{n=1}^r F_n(\tau) \over r} \right]_{r\ge2}.
\end{equation}
Namely, this simply states that the maximum of $J_\delta^2(2\tau)$
exceeds that of the average of all the $F_n$ till $n=r$,
and clearly
$\sum_{n=1}^r F_n(\tau) /r \le \max(F_n(\tau))$.
This can be proven by illustrating that the maximum of $J_\delta^2(2\tau)$ 
is larger than the maximum of any $F_n$ when $n>1$.
Using the large $x$ approximation of $J_\nu(x)$,
it is just to show that the maxima of 
$({1/ \pi\tau})\cos^2\left( 2\tau - \pi\delta/2- \pi/4 \right)$
are larger than the maxima of 
$({1/ n\pi\tau})\cos^2\left( 2n\tau - \pi\delta/2- \pi/4 \right)$, with the integer $n\ge2$.
This is obvious since the latter is enveloped by ${1/ n\pi\tau}$.
Hence in the large $\tau$ limit, we can readily
verify Eq. (\ref{mlm}) and prove that the minima of $\expval{n_R(r^*)}$ occur at $r^*=1$.
Further, the envelope of $F_1$, ${1/ \pi\tau}$, 
indicates that the minima of $\expval{n_R(r^*)}$ are also growing with $\tau$, 
as seen in Eq. (\ref{minimadelta}) and Fig. \ref{minrm}.
See also Fig. \ref{minrm100} for the numerical confirmation and the incidence time $t_\text{inc}$ as a good approximation for $\tau^*$ when $\delta$ is large.

\begin{table*}[ht]
\centering
\caption{
The numerical $\tau^*$ minimizing $\expval{n_R}$, 
the maximum of $J_\delta^2(2\tau)$, $\xi_{\delta}/2$, 
and the incidence time $t_\text{inc}=\delta/2$, for different $\delta$.
Recall that in our system 
the maximal group velocity is 2, 
hence $\delta/2$ is the incident time, for a particle initially 
at a distance $\delta$ from the detector.
}
\label{tintd}
\sisetup{detect-weight,mode=text}
\renewrobustcmd{\bfseries}{\fontseries{b}\selectfont}
\renewrobustcmd{\boldmath}{}
\newrobustcmd{\B}{\bfseries}
%
\begin{tabular}{ cccccccccccccccccccccccccccccc } 
 \Xhline{2\arrayrulewidth}
 \multicolumn{3}{c}{ ${\bm \delta}$} & 
 \multicolumn{3}{c}{\B 1} & 
 \multicolumn{3}{c}{\B 2} & 
 \multicolumn{3}{c}{\B 3} & 
 \multicolumn{3}{c}{\B 6} & 
 \multicolumn{3}{c}{\B 9} & 
 \multicolumn{3}{c}{\B 12} & 
 \multicolumn{3}{c}{\B 20} & 
 \multicolumn{3}{c}{\B 40} & 
 \multicolumn{3}{c}{\B 100} \\
 \multicolumn{3}{c}{${\bm t_\text{\bf inc}}$} & 
 \multicolumn{3}{c}{\,\,$0.500$\,\,}  & 
 \multicolumn{3}{c}{\,\,$1.000$\,\,}  & 
 \multicolumn{3}{c}{\,\,$1.500$\,\,}  & 
 \multicolumn{3}{c}{\,\,$3.000$\,\,}  &
 \multicolumn{3}{c}{\,\,$4.500$\,\,}  &
 \multicolumn{3}{c}{\,\,$6.000$\,\,}  &
 \multicolumn{3}{c}{\,\,$10.000$\,\,}  &
 \multicolumn{3}{c}{\,\,$20.000$\,\,}  &
 \multicolumn{3}{c}{\,\,$50.000$\,\,}  \\
 \multicolumn{3}{c}{${\bm \xi_{\bm \delta}/{\bf 2}}$} & 
 \multicolumn{3}{c}{0.921} & 
 \multicolumn{3}{c}{1.527} & 
 \multicolumn{3}{c}{2.101} & 
 \multicolumn{3}{c}{3.751} & 
 \multicolumn{3}{c}{5.356} & 
 \multicolumn{3}{c}{6.940} & 
 \multicolumn{3}{c}{11.110} & 
 \multicolumn{3}{c}{21.393} & 
 \multicolumn{3}{c}{51.884} \\
 %
 \multicolumn{3}{c}{${\bm \tau^*}$} & 
 \multicolumn{3}{c}{0.920} & 
 \multicolumn{3}{c}{1.527} & 
 \multicolumn{3}{c}{2.100} & 
 \multicolumn{3}{c}{3.751} & 
 \multicolumn{3}{c}{5.356} & 
 \multicolumn{3}{c}{6.940} & 
 \multicolumn{3}{c}{11.109} & 
 \multicolumn{3}{c}{21.394} & 
 \multicolumn{3}{c}{51.883} \\
 %
 \Xhline{2\arrayrulewidth}
\end{tabular}
\end{table*}

Although it is difficult to rigorously prove the dominance of $r^*=1$ 
in attaining the global minimum of $\expval{n_R(r^*)}$, 
for any value of $\delta$,
it is physically reasonable and interpretable.
%
The interpretation is that the wave packet's wave-front propagates ballistically,
as quantified by the Bessel function $J_\delta^2(2t)$,
rendering the detection probability reaching its maximum
when the detector captures the wave-front (see Fig. \ref{fnn}).
Since the detector in our setup is fixed at the target site 
and switched on/off stroboscopically, 
if we choose the first ``on'' time, at $t=\tau$, 
coinciding with the juncture wherein the wave packet's wave-front 
encompasses the target site,
this temporal alignment will maximize the probability of particle detection. 
In the event of detection failure, 
recommencing the experimental process is advantageous, 
as it offers the highest likelihood of subsequent successful particle detection.
In this sense, the global optimization requires 
the specific $\tau^*$ that maximizes $J^2_{\delta}(2\tau)$ 
and thus minimizes $\expval{n_R(1)}$,
indicating $(1,\xi_{\delta}/2)$ as the optimal set of parameters.
Notably, in large $\delta$ cases, 
a good approximation is $\tau^*=\xi_{\delta}/2\simeq \delta/2$ (Fig. \ref{minrm100}).
%




%
\section{Summary}
%
With restarts introduced to quantum hitting times,
we find novel features that expose the instability existing in the optimization of the mean hitting time.
The first feature is the presence of {\em several minima} of the mean hitting time \cite{Ruoyu2023},
rather than one unique minima, as in the classical case \cite{Gupta2022}.
This is due to the interference-induced quantum oscillations, 
indicated by the Bessel function $J_n(x)$ in Eq. (\ref{prop}) 
and the graphic illustration in Fig. \ref{fnn}.
These oscillations are found features of the solution to the Schrödinger equation, 
and the quantum first-hitting time probability $F_n$ in the absence of restarts. 
These are general aspects of quantum dynamics, 
so we expect that our results will have a wider application than the model presented here.

Then the challenge is to find the optimum of the mean hitting time under restart.
For large $\tau$ (in units $\gamma=1$),
we showed that 
$r^*(\tau)$ possesses a staircase structure of period $\pi/2$,
accompanied by plunges/rises, 
see Figs. \ref{ore0},\ref{orep1},\ref{orep2}.
We note that the $r^*$ here 
is the optimal choice of the restart step $r$ for a given sampling time $\tau$.
Furthermore, there are two symmetric patterns of staircases, 
determined by the {\em parity} of the distance between the initial and detected sites 
(in units of the lattice constant).
All those findings depict the instability existing in the quantum restart problems.
Namely, slight changes of $\tau$ lead to a large change of $\expval{n_R(r)}$ (Fig. \ref{rmht}),
optimum switching between different minima (Fig. \ref{tran}),
and change of the parity of $\delta$ causes a ``flipping''
of the staircase pattern of the optimal restart time.
We want to note that the instability is not limited in the large $\tau$ case,
and it is also found for small $\tau$ though then the staircases 
do not converge to an asymptotic limit, 
see the small $\tau$ limit of Fig. \ref{Bess0}.
Since the instability is essentially attributed to 
the oscillatory nature of the hitting time statistics,
we expect its generality when changing other control parameters. 
Is the instability we found in quantum restarts 
``robust'' in the presence of external ``disturbance'', 
such as imperfect projective measurements, or non-stroboscopic measurements?
We numerically study the latter with Monte-Carlo simulations, 
through introducing uniformly distributed noise/deviations 
to the chosen fixed sampling time.
The results indicate that the instability is quite robust 
when confronting the noise in sampling, 
in the sense that the mean hitting time around the plunge $\tau$ 
(in Fig. \ref{tran}),
still exhibits two minima up to a noise level of $20\%$, 
as shown in Fig. \ref{fig14}. 
Further research on the impact of different types of noise, 
including analytical study, is deemed worthwhile.

Another issue is the global minimum of $\expval{n_R}$ 
given a specific choice of both controlling parameters, $(r^*,\tau^*)$.
We conjecture that the global optimization occurs 
when $\tau$ is tuned to minimize the $\expval{n_R(1)}$, 
hence the restarts must be made after the first measurement.
Roughly speaking, 
when the sampling time or measurement period $\tau$ is around $\tau^*=\xi_{\delta}/2$,
which maximizes $J^2_\delta(2\tau)$,
the first-detection attempt at $\ket{\delta}$ will succeed 
with a relatively large probability. 
This leads to the optimal choice of restart time at $r=1$.
In the cases where the distance between the target and initial sites is large, i.e. $\delta\gg1$,
$\xi_{\delta}/2$ is approximately the incidence time $t_\text{inc}=\delta/2$, 
showing the ballisticity of the wave-front spreading.
Hence in the sense of detecting the quantum walker with least attempts 
(through the sharp-restart strategy),
when $\delta$ is large,
restarting after the first measurement with the highest likelihood of success, 
performed at $\tau^*\simeq\delta/2$,
is the most efficient choice.
We emphasize again that 
we optimized here $\expval{n_R}$ and not $\expval{n_R}\tau$, 
so clearly more work is needed.
Can we foresee quantum experiments checking the validity of this work?
A good platform for that aim are quantum computers,
on which on-demand qubit resets have been achieved 
with the initial purpose of optimizing quantum circuits.
The quantum walk dynamics can be mapped to a spin model 
through the Jordan-Wigner transformation \cite{Jordan1928}, 
or other methods \cite{Wang2023},
with repeated measurements implemented by the built-in single-qubit measurements 
\cite{Sabine2022}.
Similar experiments on finite systems, 
with restarts taking place after $20$ measurements,
have been successfully implemented \cite{Yin2023c}. 
This greatly enhances our confidence in verifying the theory 
presented in this paper for the optimization problem,
though finite-size effects might be important.
%

In this work we considered a quantum walk on the line, 
however this is not an essential ingredient of our results, 
in the sense that quantum instabilities for the restart problem 
can be found also for finite systems.
We also note that possibly other oscillatory dynamics or reset strategies 
\cite{Pedro2023},
beyond the quantum case, 
will exhibit similar features to what we have found here.
A key point to study the quantum instabilities is the use of sharp restart, 
which as mentioned is the optimal choice.
In our previous work \cite{Ruoyu2023}, 
we studied briefly quantum restarts with Poisson restarts, 
which did not exhibit the multiple minima found here.

\begin{acknowledgments}
The support of Israel Science Foundation's grant 1614/21 is acknowledged.
\end{acknowledgments}


%

%

\appendix

%
\section{Under restart: the first-hitting probability, guaranteed detection}
\label{appdix1}
\subsection{The probability}
In probability theory language,
the first-hitting probability
$F_n = \text{Pr}(\overline{{\cal E}_1}\overline{{\cal E}_2} 
\cdots \overline{{\cal E}_{n-1}} {\cal E}_n)$,
where ${\cal E}_k$ denotes the event of successful detection 
at the $k$th attempt,
and $\overline{{\cal E}_k}$ means the event of failing 
to detect the walker at the $k$th attempt.
Now with the deterministic restart strategy incorporated,
the new measurement protocol is as follows:
after the $r$th attempt, if the walker is not yet detected,
we start anew with the same initial state,
until the first successful detection at the $n_R$th attempt 
with $n_R=r{\cal R} + \tilde{n}$,
where ${\cal R}$ is the number of restart event and  
$\tilde{n}$ is the number of attempts till success following the last restart,
and then $1 \le \tilde{n} \le r$, ${\cal R}\ge 0$
since the restart is made just after a measurement.
Following standard statistical methods,
the first-hitting probability under restart is:
\begin{equation}\label{eq3}
\begin{aligned}
    &F_{n_R} =
                \text{Pr}
                (\underbrace{
                            \overline{{\cal E}_1} \cdots \overline{{\cal E}_r}
              \overline{{\cal E}_1} \cdots \overline{{\cal E}_r}
                            }
                            _{\text{repeat } {\cal R} \text{ times}}
                \overline{{\cal E}_1} \cdots {\cal E}_{\tilde{n}}) \\
              =&
                \left( 1 - F_1 - \cdots - F_r \right)^{\cal R} F_{\tilde{n}}
              =
                \left( 1- \sum_{k=1}^r F_k \right)^{\cal R} F_{\tilde{n}}
              .
\end{aligned}
\end{equation}
where $F_j$ are the probabilities of the first-hitting at the $j$th attempt 
in the absence of restart.
The last expression in Eq. (\ref{eq3}) is also intuitive:
the term inside the brace suggests the survival probability 
after $r$th measurement,
the power is the number of reset event,
and the term outside the brace gives 
the probability of first success after the last restart,
thus the product of the two probability is $F_{n_R}$.
Note here $\tilde{n}$ ranges from $1$ to $r$ 
(different from the remainder ranging from $0$ to $r-1$),
and ${\cal R}\ge 0$.
If $r=10$, $F_{n_R=30}$ is 
$\left( 1- \sum_{k=1}^{10} F_k \right)^{2} F_{10}$.

\subsection{Proof for the guaranteed detection}
The restarted total detection probability $P_{\rm det}$ is
%
%
\begin{equation}\label{eqA1}
\begin{aligned}
    P_{\rm det}        &= \sum_{n_R=1}^\infty F_{n_R}
                         = \sum_{r {\cal R} + \tilde{n} = 1}^\infty
                            \left( 1- \sum_{j=1}^r F_j \right)^{\cal R} F_{\tilde{n}}
                        = \sum_{{\cal R}=0}^\infty  \sum_{\tilde{n}=1}^r
                            \left( 1- \sum_{j=1}^r F_j \right)^{\cal R} F_{\tilde{n}} \\
                        &= \sum_{\tilde{n}=1}^r F_{\tilde{n}}
                            \sum_{{\cal R}=0}^\infty
                            \left( 1- \sum_{j=1}^r F_j \right)^{\cal R} 
                        = \sum_{\tilde{n}=1}^r F_{\tilde{n}}
                            \left(
                                \sum_{j=1}^r F_j
                            \right)^{-1}
                        = 1.
\end{aligned}
\end{equation}
%
%
Hence the total detection probability is one 
provided that $\sum_{j=1}^r F_j \neq 0$,
meaning that the particle will be eventually detected,
as long as there exist finite probability of detection
during one restart period.

\section{Derivation for Eq. (\ref{eq9}) in the main text}\label{appdix2}
Here we provide an alternative derivation for the mean of $n_R$ under restart.
With Eq. (\ref{eq3}) and the definition 
$\expval{n_R(r)} = \sum_{n_R=1}^{\infty} n_R F_{n_R}$, 
we obtain
\begin{equation}\label{eq5}
\begin{aligned}
    \expval{n_R(r)} =& \sum_{n_R=1}^\infty n_R \left( 1- \sum_{k=1}^r F_k \right)^{\cal R} F_{\tilde{n}}
    =   \sum_{r{\cal R} + \tilde{n}=1}^\infty
        \left( r{\cal R} + \tilde{n} \right)
        \left( 1- \sum_{k=1}^r F_k \right)^{\cal R} F_{\tilde{n}} \\
    =&  \sum_{{\cal R}=0}^\infty
        \sum_{\tilde{n}=1}^r r{\cal R}
        \left( 1- \sum_{k=1}^r F_k \right)^{\cal R} F_{\tilde{n}} +
        \sum_{{\cal R}=0}^\infty
        \sum_{\tilde{n}=1}^r
        \tilde{n}
        \left( 1- \sum_{k=1}^r F_k \right)^{\cal R} F_{\tilde{n}} \\
    =&  r \sum_{\tilde{n}=1}^r F_{\tilde{n}}
        \sum_{{\cal R}=0}^\infty
        {\cal R}
        \left( 1- \sum_{k=1}^r F_k \right)^{\cal R}  +
        \sum_{\tilde{n}=1}^r \tilde{n} F_{\tilde{n}}
        \sum_{{\cal R}=0}^\infty
        \left( 1- \sum_{k=1}^r F_k \right)^{\cal R} \\
    =& {r \left( 1- \sum_{j=1}^r F_{j} \right) \over \sum_{j=1}^r F_{j}}  +
        \underbrace{ \sum_{k=1}^r k F_{k} \left( \sum_{k=1}^r F_{k} \right)^{-1}}_{\expval{n}^r_\text{cond}}
    =  r{1-P_{\rm det}^r \over P_{\rm det}^r} + \expval{n}^r_\text{cond}.
\end{aligned}
\end{equation}
This gives the Eq. (\ref{eq9}) in the main text.

\section{Details in plotting Fig. \ref{rmtend}}
\label{appdix3}
Since Eq. (\ref{EQ3}) is valid at the transition $\tau$,
we chose
$\tau=1.369+\pi k/2$ for $r^*=6$ with $k$ an integer 
[using Eq. (\ref{**})].
Although $r^*=6$ is cutoff by $\epsilon_\text{pl}=1.353$,
we use the fact that 
$\expval{n_R(6)}(\tau=1.353)\approx\expval{n_R(6)}(\tau=1.369)=1/F_{7}(\tau=1.369)$, see Fig. \ref{tran}.
The black crosses nicely capture the $\expval{n_R(r^*)}$ 
at the transition $\tau$'s. 

\section{Transition $\tau$ in the case of $\delta=1$ and large $\tau$ limit}
\label{tran1tau}
\begin{equation}\label{trantau1}
\begin{aligned}
    %
    &\epsilon\in [0,0.218]: \expval{n_R(1)} = \min(\expval{n_R(r)}), \\
    &\epsilon\in [0.218, \epsilon_{6\to5}]: \expval{n_R(6)} < \expval{n_R(1)}, \\
    &\epsilon_{6\to5} = 0.239 = \pi/2 - 1.332, \\
    &\epsilon \in [\epsilon_{6\to5}, \epsilon_{5\to4}]:
    \expval{n_R(5)} < \expval{n_R(1)}, \\
    &\epsilon_{5\to4} = 0.291 = \pi/2 - 1.280, \\
    &\epsilon \in [\epsilon_{5\to4}, \epsilon_{4\to3}]:
    \expval{n_R(4)} < \expval{n_R(1)}, \\
    &\epsilon_{4\to3} = 0.367 = \pi/2 - 1.204, \\
    &\epsilon \in [\epsilon_{4\to3}, \epsilon_{3\to2}]:
    \expval{n_R(3)} < \expval{n_R(1)}, \\
    &\epsilon_{3\to2} = 0.490 = \pi/2 - 1.081, \\
    &\epsilon \in [\epsilon_{3\to2}, \epsilon_{2\to1}]:
    \expval{n_R(2)} < \expval{n_R(1)}, \\
    &\epsilon_{2\to1} = 0.721 = \pi/2 - 0.850. 
\end{aligned}
\end{equation}
\end{document}